\documentclass[
aps,                    
prd,                    
superscriptaddress,    
nofootinbib,            
amsmath,               %
amssymb,               %
a4paper,                
floatfix]               
{revtex4}               

\usepackage{booktabs}
\usepackage{multirow}
\usepackage[colorlinks = true,
            linkcolor = blue,
            urlcolor  = blue,
            citecolor = blue,
            anchorcolor = blue]{hyperref}

\usepackage{float}
\usepackage{amssymb}
\usepackage{graphics}
\usepackage{graphicx}
\usepackage{amsmath}
\usepackage{hyperref}
\usepackage{amsfonts}
\usepackage{subfigure}
\usepackage{xcolor}
\usepackage{cancel}
\usepackage{colordvi}
\usepackage{lipsum}
\usepackage{bbding}
\usepackage{soul}

\newcommand{\fr}{\frac}

\newcommand{\lcdm}{$\Lambda$CDM}
\newcommand{\wcdm}{$\omega$CDM}

\newcommand{\gdcdm}{$\gamma\delta$CDM}

\newcommand{\La}{\Lambda}
\newcommand{\si}{\sigma}

\newcommand{\ga}{\gamma}

\newcommand{\om}{\omega}
\newcommand{\ld}{\delta}

\newcommand{\ra}{\rightarrow}

\newcommand{\dotr}{\mbox{$\boldsymbol{\cdot}$}}

\newcommand{\be}{\begin{equation}}
\newcommand{\ee}{\end{equation}}
\newcommand{\beqa}{\begin{eqnarray}}
\newcommand{\eeqa}{\end{eqnarray}}
\newcommand{\bi}{\begin{itemize}}
\newcommand{\ei}{\end{itemize}}
\newcommand{\ben}{\begin{enumerate}}
\newcommand{\een}{\end{enumerate}}

\begin{document}

\title{Growth Index in the \gdcdm\ model}

\author{Cemsinan Deliduman}
\email{cdeliduman@gmail.com}
\affiliation{Department of Physics, Mimar Sinan Fine Arts University, Bomonti 34380, \.{I}stanbul, T\"urkiye}
\affiliation{CNRS, Universit\'e Paris Cit\'e, Laboratoire Astroparticule et Cosmologie, F-75013, Paris, France}

\author{Furkan \c{S}akir Dilsiz}
\email[Corresponding author: ]{furkansakirdilsiz@gmail.com}
\affiliation{Department of Physics, Mimar Sinan Fine Arts University, Bomonti 34380, \.{I}stanbul, T\"urkiye}

\author{Selinay Sude Binici}
\email{binici.sude@gmail.com}
\affiliation{Department of Physics, \.{I}stanbul Technical University, Maslak 34469, \.{I}stanbul, T\"urkiye}

\begin{abstract}
To better distinguish the nature of $H_0$ and $S_8$ tensions, it is necessary to separate the effects of expansion and the growth of structure. The growth index $\gamma$ was identified as the most important parameter that characterizes the growth of density fluctuations independently of the effects of cosmic expansion. In the \lcdm\ model, analyses performed with various cosmological datasets indicate that the growth index has to be larger than its theoretically predicted value. Cosmological models based on $f(R)$ gravity theories have scale-dependent growth indices, whose values are even more at odds with the growth rate data. In this work, we evaluate the growth index in the \gdcdm\ model both theoretically and numerically. Although based on $f(R)$ gravity theory, we show through several analyses with different combinations of datasets that the growth index in the \gdcdm\ model is very close in value to the \lcdm\ and the \wcdm\ models. The growth of structure is suppressed in the \gdcdm\ model, which is formulated with the extended gravitational growth framework. Upon analyzing cosmological data, we ascertain that the \gdcdm\ model is equally competitive as the \lcdm\ and the \wcdm\ models.
\end{abstract}

\maketitle

\section{Introduction \label{intro}}

Immense advances in observational cosmology since the observation of the accelerated expansion of the Universe at low redshifts \cite{SupernovaCosmologyProject:1998vns,SupernovaSearchTeam:1998fmf,SupernovaSearchTeam:2004lze,SupernovaCosmologyProject:2008ojh} brought both critical acclaim to the standard model of cosmology, the \lcdm\ model, and hints of its inner tensions. Two of these inner tensions have more prominence compared to the other problems in our view of the Cosmos. The so-called Hubble tension is more than 5$\si$ difference \cite{Freedman:2021ahq,Verde:2019ivm,CosmoVerse:2025txj} between the early universe model dependent derivation and the late universe measurements of the Hubble constant. Diverse late-time observations consistently point out a value for the Hubble constant around or more \cite{CosmoVerse:2025txj,Bhardwaj:2025kbw} than SH0ES collaboration's result, $H_0 = 73.17 \pm 0.86\ km/s/Mpc$ \cite{Riess:2021jrx,Breuval:2024lsv}, whereas early universe constraints mainly from cosmic microwave background (CMB) observations point to a lower and equally precise value, $H_0 = 67.4 \pm 0.5\ km/s/Mpc$ \cite{Planck:2018vyg}. Another prominent tension is the persistent difference between high- and low-redshift measurements of the $S_8$ parameter, which quantifies fluctuations in matter density on large scales. It is given by $S_8 = \si_8 \sqrt{\Omega_{m0}/0.3}$, where $\Omega_{m0}$ is the ratio of the present density of matter to the critical density today, and $\si_8$ is the clustering amplitude, which is the root-mean-square of matter fluctuations on scales of $8/h\ Mpc$, where $h =H_0/(100$ $\textrm{km/s/Mpc})$. The Planck Collaboration estimate from the CMB data \cite{Planck:2018vyg}, $S_8 = 0.834 \pm 0.016$, which is again dependent on the \lcdm\ model, is $2 - 3\si$ higher than the typical late-time measurement, $S_8 = 0.772 \pm 0.017$ \cite{DES:2021vln}. These tensions that much effort is given to resolve \cite{CosmoVerse:2025txj} probe different physical processes. They are two different aspects of the evolution of the Universe. Hubble tension is relevant for the expansion of the Universe and is related to the homogeneous energy density contents of the Universe. In comparison, $S_8$ is related to the growth of structure in the Universe. It quantifies the growth of large scale structure from initial fluctuations in the distribution of the non-relativistic matter. It is relevant for the evolution of the inhomogeneous component of the cosmic energy density.

To better distinguish the nature of $H_0$ and $S_8$ tensions, it is necessary to distinguish the effects of expansion and the growth of structure \cite{Linder:2005in}. The linear density fluctuations in the matter distribution are defined by density contrast, $\ld_m = \rho_m/\bar{\rho}_m -1$, where $\rho_m$ is the density of matter in the collapsing region and $\bar{\rho}_m$ is the average density of homogeneous matter in the Universe. The evolution of linear density fluctuations is determined by the linear growth factor, $D(a) = \ld_m (a) / \ld_m (a=1)$. Then, cosmic growth rate, $f(a) = d\ln D/ d\ln a$, characterizes the slowing of the linear growth factor as the Universe expands. The dependence of the cosmic growth rate on the homogeneous density of matter is empirically approximated by $f(a) \approx \Omega_m^{\, 0.6}$ \cite{Peebles:1980yev,Fry:1985,Lightman:1990,Lahav:1991wc}. Expanding the differential equation obeyed by $f(a)$ about $\Omega_m \sim 1$, Wang and Steinhardt \cite{Wang:1998gt} derived a fitting formula for the \wcdm\ model with time-dependent dark energy. For the case of the \lcdm\ model the growth index is determined by this fitting formula to be approximately $\gamma \approx 6/11 \approx 0.545$. With a series of papers \cite{Linder:2005in,Huterer:2006mva,Linder:2007hg},  Linder and his collaborators made the case that the growth index $\gamma$, defined by $\ga = \ln f(a) / \ln\Omega_m (a)$, is the most important parameter that characterizes the growth of density fluctuations independently of the effects of cosmic expansion.

Cosmological models that presume general relativity describes the gravitational interaction may have a material genesis for dark energy, which is recognized as the cause of the Universe's current accelerating expansion. If that assumption is relaxed, the plethora of alternative gravity theories source the dark energy from the geometry part of the field equations as an ``effective'' energy-momentum tensor.
As argued in \cite{Linder:2007hg} and \cite{Polarski:2007rr}, different dark energy models with general relativistic description of gravity have growth indices not too distinct from the \lcdm's one, $\gamma \sim 0.55$. In contrast, in cosmological models based on alternative or modified gravity theories, the growth index can have quite distinct values. This prediction has been supported by many works in the later literature. In \cite{Linder:2007hg} the growth index in Dvali-Gobadadze-Porrati (DGP) gravity \cite{Dvali:2000hr,Deffayet:2001pu} is calculated from the derived fitting formula to be $\gamma = 11/16 \approx 0.688$. As pointed out in \cite{Linder:2007hg} the growth index in alternative gravity theories is scale dependent and the difference from the growth index of the \lcdm\ model is given in terms of the parameterized post-Newtonian (PPN) parameter $\ga_{PPN}$ \cite{Linder:2007hg,Will:2005va,Anton:2025zer}. It is worth mentioning the growth index values in two distinct $f(R)$ gravity models to observe this scale dependence: In \cite{Gannouji:2008wt} the growth index is determined for the Starobinsky model \cite{Starobinsky:2007hu} and is found to have the value $\ga (z) = 0.399  -0.246 z$ for $\Omega_{m0} = 0.315$. In \cite{Tsujikawa:2009ku} it is argued that the value of the growth index is found to be $\ga \sim 0.4$ for many viable $f(R)$ gravity models. In contrast, in the Hu-Sawicki model \cite{Hu:2007nk,Basilakos:2013nfa} the growth index is found \cite{Basilakos:2017rgc} to be $\ga (a) = 0.753 + 0.690 (1-a)$, which is calculated on a different scale.

In the present work, we are going to calculate the growth index in the recently proposed \gdcdm\ model \cite{Deliduman:2023caa,Binici:2024smk,Dilsiz:2025ucs}. 
Our aim in this article is not to investigate either the $H_0$ or the $S_8$ tensions. The scope of this paper is to compute the growth index in the \gdcdm\ model and evaluate its potential in mitigating the S8 tension.
The \gdcdm\ cosmological model obeys $f(R)$ gravity in axially symmetric Bianchi type I background. 
In contrast to the other $f(R)$ gravity models, we neither define an effective energy-momentum tensor using the curvature terms, nor define an effective Newton constant. In this model, the contributions of cosmic energy densities to the Hubble parameter are each weighted by an equation of state parameter dependent constant. Furthermore, energy densities contribute to the Hubble parameter with a different redshift dependence, compared to what their physical nature requires. This model does not allow for a cosmological constant, but a dynamical dark energy \cite{Deliduman:2023caa}. Since the dependence of the Hubble parameter on cosmic energy densities is modified, the relation between cosmic time and redshift is also modified \cite{Binici:2024smk}. It is also demonstrated that in this model the Hubble constant is \emph{true constant} and does not run with redshift \cite{Dilsiz:2025ucs}. 
Another interesting result is that the value of the Hubble constant obtained through Bayesian analysis with various datasets is in the 1$\si$ bound of the SH0ES result \cite{Riess:2021jrx}. 

Many tensions with observations brought the cosmological community to the realization that the \lcdm\ model should be supplemented with new physics or modified \cite{CosmoVerse:2025txj,H0DN:2025lyy}. This modification could be intra-model, or it might be necessary to modify the basic assumptions. The dark energy models, such as the \wcdm\ or the CPL model \cite{Chevallier:2000qy,Linder:2002et,Linder:2003dr}, modify the dark energy sector with the aim of better explaining the data and reducing or eliminating the cosmological tensions. A complementary approach to modification is to alter the theory of gravity and utilize a modified or alternative theory of gravity \cite{Faraoni:2010pgm} instead of general relativity. In cosmological models based on modified theories of gravity, the gravitational sector of the field equations include extra ``curvature'' terms in comparison with the theory of general relativity. In the usual practice, these excess terms are moved to the matter side of the field equations and are interpreted as an effective energy-momentum tensor. This effective energy-momentum tensor is supposed to provide the necessary dark energy component with an ``exotic'' equation of state. In contrast, the \gdcdm\ model has a unique approach among the modified gravity based models. The field equations are solved exactly without assuming any artificially constructed ``effective'' energy-momentum tensor. In this model, the dark energy has material origins \cite{Deliduman:2023caa,Croker:2021duf,Farrah:2023opk}. Through many cosmological tests we aim to show that it is a competitive cosmological model to resolve the cosmological tensions and bring sound explanations to cosmological observations.

In the \gdcdm\ model, we utilize the extended gravitational growth framework \cite{Linder:2009kq,Linder:2013dga}, which is forced by the non-standard high redshift limit of the density contrast, $\delta_m \propto a^{(1-\delta)}$. Thus, the growth rate is defined by including the growth rate calibration parameter, $f_\infty$, as $f(a) = f_\infty \hat{\Omega}_m^\gamma$. We find in Section \ref{index} that the growth rate at high redshift is less than that found in models based on general relativity: $f \ra f_\infty = (1-\delta) < 1$ in the \gdcdm\ model.

In the following pages, we will show that even though the \gdcdm\ model is based on $f(R)$ gravity, the growth of structure behaves more closely to the case of \lcdm . We stress that a cosmological model that obeys modified gravity needs not to be yet another parameterization of dark energy but may bring a whole new correspondence between expansion of the Universe and its energy content. The theoretical value of the growth index, determined with Bayesian analyses with various datasets, is found to be $\ga (a=1) = 0.561$. This value is very close to the \lcdm\ value of $\ga (a=1) = 0.555$.
In the case of structure growth, the prediction of the growth index of the \gdcdm\ model is closer to the \lcdm\ and the \wcdm\ model, rather than the f(R) gravity theories.

To determine the posterior values of the cosmological parameters of the \gdcdm\ model, we perform Bayesian analyses with various datasets and their combinations. Datasets we choose to use are the $f(z)$ growth rate data \cite{Avila:2022xad}, the Dark Energy Spectroscopic Instrument Data Release 2 (DESI2) baryon acoustic oscillation (BAO) data \cite{DESI:2025zgx}, and the Dark Energy Survey Year 5 (DESY5) Type Ia supernovae (SNe Ia) data \cite{DES:2024jxu,DES:2024hip,DES:2024upw,desy5_github}. We perform two different types of analysis. In the first type, we use the theoretical expression for the growth index $\gamma$ for each model and we fit this theoretical expression together with the cosmological distances evaluated from the model to the combinations of datasets to determine the posterior values of the cosmological parameters and calculate the value of the growth index. In the second type, we treat the growth index as a free parameter in each model and perform a Bayesian analysis with this extra parameter. We find that the growth index determined from the theoretical expression is in the 1$\si$ bound of the numerically determined growth index.

We chose the datasets with the principle that we can separate the growth of structure from the effects of expansion of the Universe. $f(z)$ dataset compiled in \cite{Avila:2022xad} includes data only from direct measurements of $f(z)$, and not data obtained from $f\si_8$ measurements that eliminates the $\si_8$ dependence by using a fiducial cosmological model. An additional assumption of a primordial inhomogeneity power spectrum is required to use $f\si_8$ or $\si_8$ data, which brings additional freedom \cite{Cao:2023eja}. We also do not use data related to the value of the Hubble constant. Since the value of the Hubble constant is the point of debate and is one of the cosmological tensions, we use data in which it is possible to marginalize the value of $H_0$. As described in Section \ref{data}, this marginalization is possible in the DESI2 and DESY5 datasets.

To assess whether the \gdcdm\ model shows any potential to alleviate the $S_8$ tension, we need to constrain the values of the $\si_8$ and then the $S_8$ parameters. For that purpose, we are going to utilize two datasets: One is the 14 data points of the $\si_8(z)$ measurements compiled by \cite{Piccirilli:2024xgo}, and are presented in Table 1 of \cite{Oliveira:2025huk}. The other is the 35 data points with less correlation from the $f\si_8(z)$ dataset given in \cite{Skara:2019usd,Kazantzidis:2018rnb}. From analyses presented in section \ref{sig8} we reach the conclusion that the \gdcdm\ model is a competitive model to resolve the $S_8$ tension.
However, we stress that the resolution of the $S_8$ tension is beyond the scope of this work. We still need further analyses including the high redshift datasets to determine the relevance of the \gdcdm\ model on the $S_8$ tension.

This paper is organized as follows. In the next section, we summarize the derivation of the \gdcdm\ model in Bianchi type I background in the $f(R)$ gravity framework. Then, in Section \ref{growth}
we derive the differential equations satisfied by density contrast $\ld_m$ and growth rate $f(a)$. Afterwards, we solve the differential equation of the growth rate and calculate the growth index. In this section, we also derive a fitting formula for the growth index in the \gdcdm\ model, dependent on the cosmological parameters, $\ld$ and $\ga_e$, of the model.
In Section \ref{data} we present datasets with which we perform the Bayesian analysis, and describe the details of the method. Section \ref{results} contains presentation and discussion of the main results of our analyses: in subsection \ref{theo} we determine the growth index from the theoretical formulae, then in subsection \ref{numer} we treat the growth index as a free parameter and determine its value through Bayesian analysis, lastly in subsection \ref{models}, we compare with the growth rates calculated in other $f(R)$ gravity models by fitting them to the $f(z)$ growth rate data\footnote{Unfortunately, the same Latin letter, $f$, is used for both the function of the scalar curvature and the growth rate $f(a)$. We expect that their individual content would prevent confusion.}. In the last section, we summarize our work, examine our results, and discuss possible future works.

\section{\gdcdm\ model \label{gdcdm}}

The \gdcdm\ model is described by an exact solution of the $f(R)$ gravity field equations on an axially symmetric Bianchi type I background \cite{Deliduman:2023caa}. Due to the anisotropic nature of the background, this model also includes a contribution of anisotropic shear to the modified Friedmann equations. The details of the derivation of the model from the field equations of $f(R)$ gravity can be found in \cite{Deliduman:2023caa}.

The Hubble parameter in the \gdcdm\ model is given by\footnote{In order not to confuse with the growth index $\gamma$, we rename our model parameter to $\gamma_e$.}
\be \label{H2a}
H^2 (a) = H_0^2 \left[ \fr1{b_{\gamma_e}}\fr{\Omega_{e0}}{a^{\gamma_e-\delta}}
+ \fr1{b_3}\fr{\Omega_{m0}}{a^{3-\delta}} + \fr1{b_4}\fr{\Omega_{r0}}{a^{4-\delta}}
+ \fr1{1 -\delta}\fr{\Omega_{s0}}{a^{6 - 2\delta}} \right] \ ,
\ee
where $\Omega_{e0}, \Omega_{m0}$ and $\Omega_{r0}$ denote the dimensionless density parameters for dark energy, dust, and radiation components today, respectively. Additionally, $\Omega_{s0}$ denotes the contribution of anisotropic shear.
The coefficients $b_n$ for ($n = \gamma_e,3,4$) in (\ref{H2a}) are given by
\be \label{bn}
b_n = - \left( 1+ \delta - \fr1{2n} (n -\delta)(4+\delta) \right) .
\ee
Thus, in this model the perfect fluid components include dust ($n=3$) with vanishing pressure, radiation ($n=4$) with positive pressure and dark energy ($n=\gamma_e =3+3\omega$) with negative pressure ($-1 < \omega \leqslant -1/3$). In the $\delta = 0$ limit, all $b_n$ are unity, which is the general relativistic limit. In this limit, the \gdcdm\ model converges to the \wcdm\ model. Then the \lcdm\ model is obtained by setting $\omega = -1$ as well.

A nonzero value of the $\delta$ parameter determines both how the different perfect fluid components dissipate as the Universe expands and also how these components contribute to expansion of the Universe through the $b_n$ coefficients, which can be understood as weight factors. 
Depending on the value of $b_n$, the contribution of one component may get diminished as the contribution of another component get enhanced. For example, the \gdcdm\ model empowers a possible small density of dark energy to have a large impact on expansion of the Universe \cite{Deliduman:2023caa,Farrah:2023opk}.

According to the current paradigm, the collapse of the overdense regions starts well into the matter dominated era, and continues until today well into the dark energy dominated era. Thus, we can safely ignore the contribution of the radiation term to the Hubble parameter as affecting the collapse of overdense regions.

We also need to set a value for $\Omega_{s0}$. According to numerical analysis with various analysis packages, $\Omega_{s0}$ and $\delta$ are correlated. To lift the possible degeneracy, we can choose a value for $\Omega_{s0}$ that is not in conflict with physical expectations. This value can be guessed from an old study by Barrow \cite{Barrow:1976rda}. In that work, it is claimed that the ratio of the shear to the expansion rate is constrained by $(\sigma/\theta)_0 \lesssim 4.8\cdot 10^{-12}$. In our case $S^2(t) = 3\si^2(t)/2$ and thus $s_0^2 = 3\si_0^2 /2$. Noting also that $\theta = 3H_0$ it is found that
\be
\Omega_{s0} = \fr{s_0^2}{9 H_0^2} = \fr{3\si_0^2}{2\theta^2} \lesssim 3.5\cdot 10^{-23} \ .
\ee
This also agrees with expectations from the BBN calculations \cite{Park:2025fmu}. We can thus also ignore the contribution of the anisotropic stress to the Hubble parameter in the linear regime of collapse of the overdense regions. 

The relevant Hubble parameter is then
\be \label{H2r}
H^2 (a) = H_0^2 \left[ \fr1{b_{\gamma_e}} \fr{\Omega_{e0}}{a^{\gamma_e-\delta}}
+ \fr1{b_3} \fr{\Omega_{m0}}{a^{3-\delta}} \right] \ ,
\ee
with $\gamma_e = 3(1+\omega)$, where $\omega$ is the equation of state parameter of dark energy.

\section{Growth of structure \label{growth}}

\subsection{Density perturbation \label{fluc}}

Cosmic microwave observations paint an early universe that is almost perfectly homogeneous at the time of the last scattering. However, in this pretty picture, the fine details due to the temperature fluctuations point out the existence of overdense and underdense regions inside the overall homogeneous background. 
Consider a region of matter overdensity in this almost homogeneous Universe. We take the radius of this region to be $r$ and the matter inside is collapsing because of gravitational attraction. In the collapsing region, the perturbed density $\rho_\omega (\vec{r},t)$ of a cosmic fluid with EoS parameter $\omega$ is given by
\be
\rho_\omega = \bar{\rho}_\omega (1 + \delta_\omega) ,
\ee
where $\bar{\rho}_\omega (t)$ is the average background density of the fluid and $\delta_\omega (\vec{r},t)$ is called the density contrast or dimensionless density fluctuation\footnote{In order to prevent confusion with the cosmological parameter $\delta$ of the \gdcdm\ model, we write the density contrast with an index, like $\delta_\omega$ or $\delta_m$.}.

To derive the dynamics of the density fluctuation, its derivatives with respect to cosmic time can be evaluated utilizing the fluid equations,
\be \label{fluid}
\dot{\bar{\rho}}_\omega +3\fr{\dot{a}}a (1+w)\bar{\rho}_\omega = 0 \quad \mathrm{and} 
\quad \dot{\rho}_\omega +3\fr{\dot{r}}r (1+w)\rho_\omega = 0 ,
\ee
where $\dot{a}/a = H(a)$ is the Hubble parameter of the background universe, and $\dot{r}/r$ is the corresponding parameter in the collapsing region.

Differentiating the density contrast with respect to the cosmic time and employing the fluid equations (\ref{fluid}) \cite{Abramo:2007iu,Farsi:2022hsy,Akarsu:2025ijk} one obtains
\beqa \label{tder1}
\dot{\delta}_\omega &=& 3(1+\delta_\omega)(1+\omega)(\fr{\dot{a}}a - \fr{\dot{r}}r) , \\
\ddot{\delta}_\omega &=& 3(1+\delta_\omega)(1+\omega)(\fr{\dot{a}}a - \fr{\dot{r}}r)^{\dotr}
+ \fr{\dot{\delta}^2_\omega}{1+\delta_\omega} \ , \label{tder2}
\eeqa
where we assume that the equation of state parameters are independent of the cosmic time.

We now further assume that only the pressureless matter, with $\omega =0$, clusters. To proceed further, we need explicit expressions for $\dot{a}/a$ and $\dot{r}/r$ depending on the specific model and the assumptions regarding the collapsing region.  Combining the time derivatives (\ref{tder1},\ref{tder2}) of density fluctuation $\delta_m$ and also using the form of $H(a)$ (\ref{H2r}) of the \gdcdm\ model one reaches
\be
\ddot{\delta}_m + 2H(a)\dot{\delta}_m -\fr43 \fr{\dot{\delta}^2_m}{1+\delta_m} 
- 4\pi G \bar{\rho}_m (1+\delta_m) \Delta_m = 0\ ,
\ee
where 
\be \label{cdm}
\Delta_m = (1-\delta)\fr{a^\delta}{b_3} \left( (1+\delta_m)^{1-\ld/3} -1 \right) . 
\ee
In the limit of $\delta = 0$, $\Delta_m$ becomes $\delta_m$, which is the case of the \lcdm\ model \cite{Abramo:2007iu,Akarsu:2025ijk}. This equation differs from the one of the \lcdm\ model by the existence of the $\Delta_m$ factor and the form of $H(a)$ (\ref{H2a}).

Since the Hubble parameter is given as a function of the scale factor $a$, we change the independent variable from cosmic time $t$ to the scale factor. We then obtain
\be \label{nl1}
\delta^{\prime\prime}_m + \left( \fr3a + \fr{H^{\prime}}H \right) \delta^{\prime}_m 
- \fr43 \fr{(\delta^{\prime}_m)^2}{1+\delta_m} - \fr3{2a^2} \Omega_m (1+\delta_m) \Delta_m = 0 ,
\ee
where $\Omega_m (a) = \Omega_{m0}H_0^2/a^3 H^2(a)$, with 
$\Omega_{m0} = 8\pi G \bar{\rho}_{m0}/3H_0^2$ the dimensionless matter density parameter today, and $H_0$ is the Hubble constant.

\subsection{Growth rate \label{rate}}

The linear growth factor $D(a)$ evolves the linear density fluctuation with scale factor as given by
\be \label{grfunc}
\delta_m (a) = D(a) \delta_m (a=1) ,
\ee
where $a=1$ corresponds to the present value of the scale factor. One further defines the cosmic growth rate $f(a)$ as
\be \label{grlgf}
f(a) = \fr{d\ln D}{d\ln a} = a \fr{\delta^{\prime}_m}{\delta_m} .
\ee
which characterizes the slowing of the linear growth factor as the Universe expands.
Utilizing the definition of $f(a)$ in terms of $\delta^{\prime}_m /\delta_m$, the nonlinear evolution equation for the matter perturbation (\ref{nl1}) can be rewritten in the linear regime in terms of $f(a)$ and $f^{\prime}(a)=df/da$ as
\be \label{gre1}
f^{\prime} + \left( \fr2a + \fr{H^{\prime}}H \right) f 
+ \fr {f^2}a - \fr3{2a} \Omega_m (a) \fr{\Delta_m}{\delta_m} = 0\ .
\ee
This equation differs again from the \lcdm\ counterpart \cite{Akarsu:2025ijk} in the form of the Hubble parameter (\ref{H2a}) and the linearized form of $\Delta_m = \delta_m(1-\delta)(3-\delta) a^\delta/(3b_3)$ (\ref{cdm}) multiplying the last term.

Using the expression for the Hubble parameter in the \gdcdm\ model (\ref{H2r}) we find that
\be
\fr{H^{\prime}(a)}{H(a)} = -\fr{3-\delta}{2a} \left[ 1 + \fr{3\omega}{3-\ld}
\left( 1 - \hat{\Omega}_m (a) \right) \right] \ ,
\ee
where we defined
\be \label{hatO}
\hat{\Omega}_m (a) = \Omega_m (a) \fr{a^\delta}{b_3} 
= \fr{\Omega_{m0}/a^3}{H^2(a)/H_0^2} \fr{a^\delta}{b_3} \ .
\ee
Then the differential equation satisfied by the growth rate $f(a)$ in the \gdcdm\ model takes the form of
\be \label{gre2}
\fr{df}{d\ln a} + f^2 + \fr12 \left[ 1+\ld - 3\omega \left( 1 - \hat{\Omega}_m (a) \right) \right] f 
- \fr12 (1-\delta)(3-\delta) \hat{\Omega}_m (a) = 0\ .
\ee
Since at $a=1$ we can solve $\Omega_{e0}$ in terms of $\Omega_{m0}$ as 
$\Omega_{e0}/b_{\gamma_e} = 1 - \Omega_{m0}/b_3$ utilizing the definition of the Hubble parameter (\ref{H2r}), the differential equation (\ref{gre2}) depends on just three cosmological parameters: $\Omega_{m0}, \omega$ and $\delta$.

By a further change of variable, we rewrite equation (\ref{gre2}) as
\be \label{greO}
3\omega\hat{\Omega}_m (1-\hat{\Omega}_m)\fr{df}{d\hat{\Omega}_m} 
+ f^2 + \fr12 \left[ 1+\ld - 3\omega \left( 1 - \hat{\Omega}_m \right) \right] f 
- \fr12 (1-\delta)(3-\delta) \hat{\Omega}_m = 0\ ,
\ee
which now depends on the variable $\hat{\Omega}_m$.

This equation can be solved in terms of hypergeometric functions. The form of the function $f(\hat{\Omega}_m)$ around the singularity $\hat{\Omega}_m = 1$ is found to be
\be \label{fQ}
f(\hat{\Omega}_m) = (1-\delta)
\frac{\,_2F_1 \left[ 1+A\ , \frac{1}{2}-B\ ; 1+A-B\ ; a^{-3\omega} \Big( 1- \fr{b_3}{\Omega_{m0}}\Big) \right]}
{\,_2F_1 \left[ A\ , \frac{1}{2}-B\ ; 1+A-B\ ; a^{-3\omega} \Big( 1- \fr{b_3}{\Omega_{m0}}\Big) \right]} \ ,
\ee
where $A = (\delta-1)/3 \omega$ and $B = (3 -\delta)/6 \omega$.

From this solution, it is possible to read the growing solution for the density contrast in the \gdcdm\ model. It can be expressed as
\be \label{dmgs}
\delta_m (a) = a^{(1-\delta)} \,_2F_1 \left[ A\ , \frac{1}{2}-B\ ; 1+A-B\ ; a^{-3\omega} \Big( 1- \fr{b_3}{\Omega_{m0}}\Big) \right] \ .
\ee
In the limit of general relativity, $\delta \ra 0$, this growing solution for $\delta_m (a)$ becomes the same as the growing solution in the \wcdm\ model \cite{Silveira:1994yq,Lee:2009gb,Nesseris:2017vor,Velasquez-Toribio:2020xyf}.

The high redshift limits of the density contrast and the growth rate,  
\be \label{hzlim}
\delta_m \propto a^{(1-\delta)} \quad \mathrm{and} \quad f \propto (1-\delta),
\ee
respectively suggest that the formulation of structure growth in the \gdcdm\ model necessitates the use of the extended gravitational growth framework \cite{Linder:2009kq,Linder:2013dga} with two parameters, the growth index and the growth rate calibration, as we describe in the next subsection.

\subsection{Growth index and growth rate calibration \label{index}}

The cosmic growth rate, defined by
\be
f(a) = \fr{d\ln D}{d\ln a}\ ,
\ee
characterizes the slowing of the linear growth factor as the Universe expands. The dependence of the cosmic growth rate on the homogeneous density of matter is empirically approximated by $f(a) \approx \Omega_m^{\, 0.6}$ \cite{Peebles:1980yev,Fry:1985,Lightman:1990,Lahav:1991wc}. With a series of papers \cite{Linder:2005in,Huterer:2006mva,Linder:2007hg},  Linder and his collaborators made the case that a good approximation for $f(a)$ is given by
\be
f (a) \simeq \Omega_m^\gamma (a) ,
\ee
where $\gamma(a)$ is called the growth index. 
Although different possibilities are considered in \cite{Linder:2007hg}, it is possible to define the growth index simply by 
\be \label{gri}
\ga = \fr{\ln f(a)}{\ln\Omega_m (a)}\ ,
\ee
which is the most important parameter that characterizes the growth of density fluctuations independently of the effects of cosmic expansion.

As explained in detail in \cite{Deliduman:2023caa,Binici:2024smk,Dilsiz:2025ucs} and in Section \ref{gdcdm}, in the \gdcdm\ model the contribution of the pressureless matter to the expansion of the Universe is not $\Omega_{m0}/a^3$, but rather $(\Omega_{m0}/b_3)/a^{(3-\delta)}$. In addition, we note that the growth rate $f(a)$ (\ref{fQ}) approaches $(1-\delta)$ at the high redshift limit (\ref{hzlim}). 
Therefore, we define the growth rate in terms of the growth index by
\be \label{fgex}
f(a) = (1-\delta) \hat{\Omega}_m^\gamma (a) ,
\ee
where $\hat{\Omega}_m$ is defined by equation (\ref{hatO}). Then the growth index is obtained by
\be \label{hgri}
\ga = \fr{\ln (f(a)/(1-\delta))}{\ln\hat{\Omega}_m (a)}\ .
\ee

When initial conditions result in something other than $\delta_m \propto a$, as is the case for the \gdcdm\ model (\ref{hzlim}), the parametrization of the growth must be redone \cite{Linder:2009kq,Linder:2013dga}. The addition of the growth rate calibration parameter, $f_\infty$, establishes just that. The multiplicative form of the growth rate, given by
\be \label{finf}
f(a) = f_\infty \hat{\Omega}_m^\gamma
\ee
is forced in some cosmological models by the nonstandard behavior of structure growth in the high redshift matter-dominated era, as hinted by the nonstandard limit of the density contrast, $\delta_m \propto a^{f_\infty}$ \cite{Linder:2013dga}. In essence, $f_\infty$ calibrates the growth behavior and measures the deviation from the general relativistic expectation. The difference of $f_\infty$ from unity can be determined from the  observations of redshift space distortions \cite{Linder:2013dga,Huterer:2013xky}.

In the \gdcdm\ model we find that $f_\infty = (1-\delta) < 1$ (\ref{fQ}), which could be due to the enhanced effect of expansion of the Universe on the structure growth in the \gdcdm\ model, similar to the early dark energy (EDE) model \cite{Linder:2009kq}. We note that in the \gdcdm\ model, as in the EDE model, the sound horizon at baryon drag is less than the one calculated in the \lcdm\ model: $r_d = 137.1$ Mpc in the \gdcdm\ model versus $r_d =  147.1$ Mpc \cite{Planck:2018vyg} in the \lcdm\ model. A lower sound horizon value indicates an enhancement of expansion rate in the early Universe and consequently a suppression of growth in comparison to models based on general relativity. We discuss the effect of this suppression of the structure growth on the integrated Sachs-Wolfe (ISW) effect in Section \ref{models}.

Equation (\ref{hgri}) is the theoretical expression that we use to evaluate the growth index in the \gdcdm\ model, after fitting (section \ref{theo}) the growth rate function (\ref{fQ}) to the f(z) growth rate data (Section \ref{fz}). Similarly, equation (\ref{gri}) is the theoretical expression that we use to evaluate the growth index in the \wcdm\ and the \lcdm\ models, again after Bayesian fit (section \ref{theo}) of the growth rate functions,
\be \label{fw}
f(\Omega_m) =
\frac{_2F_1 \left[ 1-\fr1{3\om}\ , \fr12 -\fr1{2\om}\ ; 1-\fr5{6\om}\ ; a^{-3w}(1- \Omega_{m0}^{-1}) \right]}
{_2F_1 \left[ -\fr1{3\om}\ , \fr12 -\fr1{2\om}\ ; 1-\fr5{6\om}\ ; a^{-3w}(1- \Omega_{m0}^{-1}) \right]} \ ,
\ee
and
\be \label{fL}
f(\Omega_m) =
\frac{_2F_1 \left[ 4/3\ , 1\ ; 11/6\ ; a^{3}(1- \Omega_{m0}^{-1}) \right]}
{_2F_1 \left[ 1/3\ , 1\ ; 11/6\ ; a^{3}(1- \Omega_{m0}^{-1}) \right]} \ ,
\ee
given for the \wcdm\ \cite{Nesseris:2017vor} and \lcdm\ models, respectively. The growth rate for the \lcdm\ model (\ref{fL}) is obtained from the \wcdm\ model's one (\ref{fw}) by setting $\omega=-1$. Similarly, the growth rate for the \wcdm\ model (\ref{fw}) can be obtained from the \gdcdm\ model's one (\ref{fQ}) by setting $\delta=0$.

\subsection{An approximate formula for the growth index in the \gdcdm\ model \label{fitting}}

For dark energy models in the matter-dominated era, an approximate theoretical formula for $\gamma$ can be obtained by expanding the related differential equation obeyed by $f(a)$ about $\Omega_m \sim 1$. In \cite{Wang:1998gt}, Wang and Steinhardt derived such a theoretical formula, given by
\be \label{gamma}
\ga = \ga_0 + \ga_1 (1-\Omega_m) + \mathcal{O}((1-\Omega_m)^2)
\ee
with
\be \label{gwcdm}
\ga_0 = \fr{3 -3\om}{5 -6\om} \quad \mathrm{and} \quad
\ga_1 = \fr12 \fr{(3 -3\om)(2 -3\om)}{(5 -6\om)^3}
\ee
for the \wcdm\ model with time-dependent dark energy. For the case of \lcdm\ model the growth index is determined by this fitting formula to be approximately $\gamma \approx 0.555$ for $\Omega_{m0} \approx 0.3$. 

In the \gdcdm\ model we define the growth index as 
\be \label{fgOa}
f(a) = (1-\delta)\cdot (\hat{\Omega}_m)^{\gamma(a)}\ .
\ee 
Expanding the differential equation (\ref{greO}) for $f(\hat{\Omega}_m)$ about $\hat{\Omega}_m \approx 1$,
we then obtain $\gamma (\hat{\Omega}_m)$ to have the form
\be \label{gaa}
\ga (a) = \ga_0 + \ga_1 (1-\hat{\Omega}_m) + \mathcal{O}((1-\hat{\Omega}_m)^2)
\ee
with
\beqa \label{g0}
\ga_0 &=& \fr{3-\delta -3\om}{5 -3\delta -6\om}\ , \\
\ga_1 &=& \fr{(1+\delta)}2
\fr{(3 -\delta -3\om)(2 -2\delta -3\om)}{(5 -3\delta -6\om)^3}. \label{g1}
\eeqa
This is our theoretical fit formula for the growth index $\gamma$. In the $\ld \ra 0$ limit, $\ga$ approaches its \wcdm\ expression (\ref{gamma},\ref{gwcdm}), first given in \cite{Wang:1998gt}. 

We also fitted the from of the growth rate given in (\ref{fgOa}) to the $f(z)$ growth rate data, and calculated the growth index by (\ref{gaa}-\ref{g1}) with the posterior values of the cosmological parameters. The results were extremely close to the value of the growth index calculated by equation (\ref{hgri}), after fitting the functional form of the growth rate (\ref{fQ}) to the $f(z)$ growth rate data. Thus, in Section \ref{theo} we only share the results of the analyses performed with the functional form of $f(a)$ (\ref{fQ}).

\section{Data and method \label{data}} 

To determine the posterior values of the cosmological parameters of the \gdcdm\ model, we perform Bayesian analyses with various datasets and their combinations. Datasets we choose to use are the $f(z)$ growth rate data \cite{Avila:2022xad}, the Dark Energy Spectroscopic Instrument Data Release 2 (DESI2) baryon acoustic oscillation (BAO) data \cite{DESI:2025zgx}, and the Dark Energy Survey Year 5 (DESY5) Type Ia supernovae (SNe Ia) data \cite{DES:2024jxu,DES:2024hip,DES:2024upw,desy5_github}. We perform two different types of analysis. In the first type, we use the theoretical expression for the growth index $\gamma$ for each model and we fit this theoretical expression together with the cosmological distances evaluated from the model to the combinations of datasets to determine the posterior values of the cosmological parameters and compute the value of the growth index. In the second type, we treat the growth index as a free parameter in each model and perform Bayesian analyses with this extra parameter. We find that the growth index determined from the theoretical expression is in the 1$\si$ bound of the numerically determined growth index.

To assess whether the \gdcdm\ model shows any potential to alleviate the $S_8$ tension, we also need to constrain the $\si_8$ parameter in this model. For that purpose, we are going to utilize two datasets: One is the 14 data points of the $\si_8(z)$ measurements compiled by \cite{Piccirilli:2024xgo}, and are presented in Table 1 of \cite{Oliveira:2025huk}. The other is the 35 data points with less correlation from the $f\si_8(z)$ dataset given in \cite{Skara:2019usd,Kazantzidis:2018rnb}. However, we stress that we still need more analyzes, including high redshift datasets, to determine the relevance of the \gdcdm\ model to the $S_8$ tension.

\subsection{Growth rate data \label{fz}} 

We use measurements of the growth rate $f(z)$ which are shown in Table 1 of \cite{Avila:2022xad}. The authors selected f(z) data from uncorrelated redshift bins when considering the same cosmological tracer and data from possibly correlated bins when different tracers were involved. They only included data from direct measurements of $f(z)$, excluding measurements such as $f\sigma_8$ that rely on cosmological assumptions, and when the same survey provided multiple data releases, they used the most recent one. For completeness, we present 11 growth rate $f(z)$ data points in the redshift range $0.013\le z \le 1.40$ in Table \ref{fzdata}. 

The chi-squared function for the growth rate $f(z)$ dataset is given by
 \be
 \chi^2 _{f(z)} = \sum _{i=1} ^{11} \fr{\left[f_{\text{data}}(z_i)-f_{\text{model}}(z_i)\right]^2}{\sigma^2 _i}.
 \ee
where $f_{\text{model}}(z_i)$ is the theoretical growth rate and $f_{\text{data}}(z_i)$ and $\sigma_i$ are the measurements of the growth rate and its associated uncertainty, as given in Table \ref{fzdata}.

\subsection{$\sigma_8 (z)$ data \label{ssigma8}} 

We use 14 data points for the amplitude of matter fluctuations, $\si_8(z)$, as compiled in \cite{Piccirilli:2024xgo,Oliveira:2025huk}. We are not using the first data point in Table 1 of \cite{Oliveira:2025huk}, since it is obtained assuming a specific value for the Hubble constant. For completeness, we present 14  $\sigma_8 (z)$  data points in the redshift range $0.24\le z \le 3.80$ in Table \ref{s8data}.

The chi-squared function for the $\sigma_8 (z)$ dataset is given by
 \be
 \chi^2 _{\sigma_8} = \sum _{i=1} ^{14} \fr{\left[\si_8^{\text{data}}(z_i)-\si_8^{\text{model}}(z_i)\right]^2}{\sigma^2 _i}.
 \ee
where $\si_8^{\text{model}}(z_i)$ is the theoretical amplitude of matter fluctuations and $\si_8^{\text{data}}(z_i)$ and $\sigma_i$ are the measurements of the amplitude of matter fluctuations and associated uncertainties, as given in Table \ref{s8data}.

As already stated in the Introduction, we do not want to have an assumption of a primordial inhomogeneity power spectrum in this work. Thus we evaluate the amplitude of matter fluctuations, $\si_8(z)$, from \cite{Nesseris:2017vor}
\be
\si_8 (z) = \si_8 \frac{\delta_m (z)}{\delta_m (0)},
\ee
where $\delta_m (z)$ can be obtained from (\ref{dmgs}) after substitution $a=1/(1+z)$.

\subsection{$f\sigma_8 (z)$ data \label{fssigma8}} 

We use a subset of $f\sigma_8 (z)$ data compiled in \cite{Skara:2019usd,Kazantzidis:2018rnb}. This subset includes 35 data points with less correlation, which are indicated in bold font in Table VI of appendix B of \cite{Skara:2019usd}. 
For the $f\sigma_8$ dataset, the $\chi^2$ function is defined by
\begin{equation}
\chi^2_{f\sigma_8}=\Delta_{f\sigma_8} C^{-1}_{f\sigma_8} \Delta_{f\sigma_8} ^T
\end{equation}
where $C^{-1}_{f\sigma_8}$ is the inverse covariance matrix, and $\Delta_{f\sigma_8}$ is defined \cite{Skara:2019usd,Kazantzidis:2018rnb} by
\begin{equation}
\Delta_{f\sigma_{8,i}}= f\sigma^{obs}_{8,i} - \frac{f\sigma^{model}_8 (z_i,\theta) }{q(z_i,\theta,\Omega^{fid}_{0m})}
\end{equation}
with $\theta$ representing the model parameters. $q$ is the fiducial Alcock-Paczynski correction factor \cite{Nesseris:2017vor,Skara:2019usd,Kazantzidis:2018rnb,Macaulay:2013swa}, given by 
\begin{equation}
q(z_i,\theta,\Omega^{fid}_{0m}) = \frac{H(z_i,\theta) d_A (z_i,\theta)}{H^{fid}(z_i) d_A^{fid} (z_i)}
\end{equation}
where $H$ is the Hubble parameter and $d_A$ is the angular diameter distance. The superscript \textit{``fid''} is an abbreviation for fiducial cosmology. We present the relevant data points and the Alcock-Paczynsk correction factor $q$ for the \gdcdm\ model, the \lcdm\ model with $\Omega_{m0} = 0.299$, and the \lcdm\ model with $\Omega_{m0} = 0.315$ in Table \ref{fsigma8}.

The covariance matrix, $C_{f\sigma_8}$, is diagonal except for the WiggleZ and SDSS IV components.
For the WiggleZ data, it is given by \cite{Kazantzidis:2018rnb,Akarsu:2025ijk},
\begin{equation}
C_{\text{WiggleZ}} = 
\begin{pmatrix}
0.00640 & 0.002570 & 0.0 \\
0.00257 & 0.003969 & 0.002540 \\
0.0 & 0.002540 & 0.005184   
\end{pmatrix}\ ,
\end{equation}
and for the SDSS IV data, the covariance matrix is \cite{eBOSS:2018yfg,Sagredo:2018ahx,daSilva:2020mvk,Akarsu:2025ijk}
\begin{equation}
C_{\text{SDSS}} = 
\begin{pmatrix}
0.0310 & 0.0089 & 0.0033 & -0.0002\\
0.0089 & 0.0098 & 0.0044 & 0.0008\\
0.0033 & 0.0044 & 0.0049 & 0.0035\\ 
-0.0002 & 0.0008 & 0.0035 & 0.0112
\end{pmatrix}\ .
\end{equation}

\subsection{DESI BAO data \label{desi}} 

We use the Dark Energy Spectroscopic Instrument (DESI) Data Release 2 (DR2) baryon acoustic oscillation (BAO) data \cite{DESI:2025zgx}. The DESI DR2 dataset consists of $D_M /r_d, D_H /r_d, D_V /r_d$ measurements in the redshift range $0.295\le z \le 2.330$ where  $D_M$ is the comoving distance, $D_H$ is the Hubble distance, $D_V$ is the angle-averaged distance and $r_d$ is the sound horizon at the baryon drag epoch. All these data and their uncertainties can be found in Table IV of \cite{DESI:2025zgx}. The chi-squared function for the DESI DR2 BAO dataset is given by 
 \be
 \chi^2 _{DESI} =  \left[\nu_{data}(z_i)-\nu_{model}(z_i) \right] C_{DESI} ^{-1}  \left[\nu_{data}(z_i)-\nu_{model}(z_i) \right]^T
 \ee
where $\nu_{data}$ is the vector of $D_V/r_d$, $D_M/r_d$ and $D_H/r_d$ measurements from the DESI DR2 dataset and $C^{-1}_{DESI} $ is the inverse covariance matrix. We use the $D_V/r_d$ measurement in the first bin, along with 12 $D_M/r_d$  and $D_H/r_d$ measurements from 6 redshift bins. Due to the correlations between $D_M/r_d$ and  $D_H/r_d$ within the same redshift bins, we use the covariance matrix in our analysis. The covariance matrix is a block-diagonal matrix, where each block corresponds to a redshift bin and is given by
\be
C_{\text{block}} = 
\begin{pmatrix}
\sigma_M ^2 & r_{M,H}\sigma_M \sigma_H\\
r_{M,H}\sigma_M \sigma_H& \sigma_H ^2
\end{pmatrix}
\ee
where $\sigma_M$ and $\sigma_H$ are the uncertainties of $D_M /r_d$ and $D_H /r_d$, respectively, and $r_{M,H}$ is the cross-correlation coefficient between $D_M /r_d$ and $D_H /r_d$ in the same redshift bin \cite{DESI:2025zgx}. For the first redshift bin, the $D_V/r_d$ measurement is uncorrelated, so its uncertainty is added as a separate diagonal element in the covariance matrix.

\subsection{DES supernovae data \label{des}} 

We use the Dark Energy Survey Year 5 (DESY5) Type Ia supernovae (SNe Ia) dataset \cite{DES:2024jxu,DES:2024hip,DES:2024upw}. The DESY5 sample consists of 1635 photometrically classified SNe Ia and 194 external low-redshift SNe Ia. Thus, the total number of SNe Ia data used is 1829, covering a redshift range $0.025 < z < 1.13$ . The data sample can be found in the GitHub repository \cite{desy5_github}.
 To compute the difference between data and theory, we calculate the distance modulus residual, which is given by $\Delta\mu_{i} = \mu_{\text{data},i} - \mu_{\text{model}}(z_i)$. Here, $\mu_{\text{data},i}$ is the observed distance modulus of the $i$-th SNe Ia, and $\mu_{\text{model}}(z_i)$ is the theoretical distance modulus at redshift $z_i$. We use the covariance matrix $C_{ij}$ to incorporate the uncertainties of the SNe Ia data. The corresponding chi-squared function is given by 
\begin{equation}
\chi^2_{DES} =  \Delta\mu_i C^{-1}_{ij} \Delta\mu_j^T,
\end{equation}
where the inverse covariance matrix $C^{-1}_{ij}$ contains both statistical and systematic errors. We use the DESY5 likelihood and covariance matrix as provided in \cite{desy5_github}.

\subsection{Bayesian analysis \label{Bayes}}

To constrain the parameters of the cosmological models, we use Bayesian inference \cite{Padilla:2019mgi,Hogg:2010}. The likelihood function for the model is defined by
\be \label{like}
\mathcal{L}\left(\mathcal{I}|\mathcal{\theta},\mathcal{M}\right) \propto \exp \left[ -\fr{1}{2} \chi^2 \left(\mathcal{I}|\mathcal{\theta},\mathcal{M}\right)\right]
\ee
where the function $ \chi^2 \left(\mathcal{I}|\mathcal{\theta},\mathcal{M}\right)$ is the chi-squared function and $\mathcal{M}$, $\mathcal{\theta}$ and $\mathcal{I}$ represent the model, model parameters, and datasets, respectively. The minimum value of the chi-squared function is equivalent to the maximum value of the likelihood function. The total chi-squared function is
\be \label{chis2}
\chi^2 _{total} = \chi^2 _{f(z)} +\chi^2 _{DESI} + \chi^2 _{DES}.
\ee
in the case of analysis with the combination of datasets.

We use the Markov Chain Monte Carlo (MCMC) sampler emcee\footnote{\url{https://emcee.readthedocs.io/en/stable/}} \cite{Foreman-Mackey:2012any} to obtain parameter constraints and generate contour plots using GetDist \cite{Lewis:2019xzd}.
We perform our analyses with three different cosmological models: the $\gamma\delta$CDM model, the $\Lambda$CDM model and the $\omega$CDM model. We analyze all models first with only the $f(z)$ dataset, then with a combination of the $f(z)$ and DESI DR2 datasets, and lastly with a combination of the $f(z)$, DESI DR2, and DESY5 datasets. We perform two different types of analysis. In the first type, we use the theoretical expression for the growth index $\gamma$ for each model and we fit this theoretical expression together with the cosmological distances evaluated from the model to the combinations of datasets to determine the posterior values of the cosmological parameters and calculate the value of the growth index. In the second type, we treat the growth index as a free parameter in each model and perform Bayesian analysis with this extra parameter. We find that the growth index determined from the theoretical expression is in the 1$\si$ bound of the numerically determined growth index.

Uniform prior distributions are chosen for the parameters of the $\gamma\delta$CDM model, the $\Lambda$CDM model, and the $\omega$CDM model. The parameters of the $\gamma\delta$CDM model have the following ranges: $0.2 < \Omega_{m0} < 0.4$, $-0.9997 < \omega < -1/3$ and $0.0 < \delta < 0.772$. The parameter of the $\Lambda$CDM model has the following range: $0.2 < \Omega_{m0} < 0.4$. The parameters of the $\omega$CDM model have the following ranges: $0.2 < \Omega_{m0} < 0.4$, $-2.0 < \omega < 0.5$. The prior range for the dark energy equation of state parameter in the $\omega$CDM model is the same as in the DESY5 \cite{desy5_github} data file.

BAO measurements do not directly measure absolute distances. Consequently, they cannot determine $H_0$ and $r_d$ separately; only the combined term $H_0 r_d $ can be measured. Before starting the growth rate analyses, we performed the DESI DR2 analyses for each cosmological model. In these analyses, we treated $hr_d$ as a free parameter constrained within the range $60 < hr_d < 120$ (Mpc), where $h =H_0/(100$ $\textrm{km/s/Mpc})$. We find $hr_d=99.11_{-1.36}^{+1.26}$ Mpc for the $\gamma\delta$CDM model.  Then, confirming with \cite{DESI:2025zgx}, we obtain $hr_d=101.54_{-0.72}^{+0.72}$ Mpc for the $\Lambda$CDM model and $hr_d=99.86_{-1.64}^{+1.72}$ Mpc for the $\omega$CDM model. We fixed these values when we used the DESI DR2 dataset in the growth rate analyses.

In the DES supernova data, the Hubble constant $H_0$ and the absolute magnitude $M$ are degenerate, thus they are treated as a single parameter $\mathcal{M} = M + 5\log_{10} (c/H_0)$. This parameter is analytically marginalized in likelihood, which means that the value of $H_0$ has no impact on the analysis. Therefore, the DES supernova dataset cannot be used for direct $H_0$ measurements.
Since the DES dataset is insensitive to $H_0$ and we keep $h r_d$ fixed when we use the DESI DR2 dataset, all of our analyzes are independent of the value of the Hubble constant.

\section{Results and Discussion \label{results}} 

This section contains a presentation and discussion of the main results of our analyses: in Subsection \ref{theo} we determine the growth index from the theoretical formulae given in Sections \ref{rate} and \ref{index}, then in Subsection \ref{numer} we treat the growth index as a free parameter and determine its value through Bayesian analysis, and lastly in Subsection \ref{models}, we compare with the growth rates calculated in other $f(R)$ gravity models by fitting them to the $f(z)$ growth rate data.

\subsection{Analyses with theoretical formulae for $\gamma$ \label{theo}}

Theoretical formulae for the growth rate as a function of redshift are given in equations (\ref{fQ}), (\ref{fw}) and (\ref{fL}) for the \gdcdm , the \wcdm , and the \lcdm\ models, respectively. We fit these cosmological models to combinations of $f(z)$ growth rate data \cite{Avila:2022xad}, the DESI2 BAO data \cite{DESI:2025zgx} and the DESY5 SNe Ia data \cite{DES:2024jxu,DES:2024hip,DES:2024upw,desy5_github}. We then evaluate the growth index from the formulae given in equations (\ref{gri}) and (\ref{hgri}). 
The posterior values of the cosmological parameters of each model and the calculated values of the growth indices are presented in Table \ref{theor}. In the same table, we also share the difference of the specific model's chi-squared function from the \lcdm\ one's. In the case of analysis only with the $f(z)$ growth rate data, there are not many data points and therefore the posterior $1\si$ confidence regions are quite large. However, the $f(z)$ growth rate data combined with the DESI2 BAO data and the DESY5 SNe Ia data constrain the models much better, with smaller $1\si$ confidence regions.

\begin{table}[hbt!]
\centering
\def\arraystretch{1.5}
\resizebox{\textwidth}{!}{%
\begin{tabular}{|l|c|c|c|c|c|c|c|c|}
\hline 
\multicolumn{1}{|l|}{} & \multicolumn{1}{|l|}{} & \multicolumn{4}{c|}{\textbf{Parameters}} & \multicolumn{3}{c|}{\textbf{Criteria}} \\
\hline 
\hline 
\textbf{Dataset} & \textbf{Models} & $\ga (a=1)$ & $\Omega_{m0}$ & $\omega$ & $\delta$ 
& $\Delta\chi^2$ & $\Delta$AIC & $\Delta$BIC \\
\hline 
\hline
 & \gdcdm & $0.567$ & $0.347_{-0.045}^{+0.036}$ 
& $-0.751_{-0.127}^{+0.175}$ & $0.069_{-0.049}^{+0.071}$ & $-0.37$ & 3.63 & 4.43  \\
f(z) & \wcdm & $0.561$ & $0.303_{-0.055}^{+0.056}$ & $-0.827_{-0.291}^{+0.260}$ & | & $-0.37$ & 1.63 & 2.03 \\
 & \lcdm & $0.555$ & $0.274_{-0.031}^{+0.034}$ & $-1$ & | & $0$ & 0 & 0 \\
\hline
 & \gdcdm & $0.561$ & $0.300_{-0.008}^{+0.007}$ 
& $-0.896_{-0.028}^{+0.029}$ & $0.042_{-0.029}^{+0.038}$ & $-1.89$ & 2.11 & 4.47  \\
f(z)+DESI2 & \wcdm & $0.558$ & $0.296_{-0.008}^{+0.008}$ & $-0.913_{-0.029}^{+0.028}$ & | & $-1.95$ & 0.05 & 1.23 \\
 & \lcdm & $0.555$ & $0.297_{-0.003}^{+0.003}$ & $-1$ & | & $0$ & 0 & 0 \\
\hline
\multirow{2}{*}{\multirow{2}{*}{\shortstack{\(f(z)\) + DESI2 \\ + DESY5}}} 
& \gdcdm & $0.561$ & $0.299_{-0.008}^{+0.007}$ 
& $-0.890_{-0.027}^{+0.028}$ & $0.040_{-0.028}^{+0.040}$ & $-13.97$ & $-9.97$ & 1.08 \\
& \wcdm & $0.558$ & $0.294_{-0.007}^{+0.007}$ &  $-0.900_{-0.027}^{+0.026}$ & | & $-12.57$ & $-10.57$ & $-5.05$ \\
 & \lcdm & $0.555$ & $0.299_{-0.003}^{+0.003}$ & $-1$ & | & $0$ & 0 & 0 \\
\hline
\hline
\end{tabular} 
}
\caption{The marginalized posterior values of the cosmological parameters, with 1$\sigma$ confidence regions, of each model obtained after data analyses with three distinct datasets. The theoretical values of the growth indices are evaluated from equations (\ref{gri}) and (\ref{hgri}). $\Delta\chi^2$ compares each model's chi-squared function to the \lcdm's chi-squared function: $\Delta\chi^2 = \chi^2_{model} - \chi^2_{\La\mathrm{CDM}}$. Differences of information criterion values are calculated with $\Delta \mathrm{AIC} = \mathrm{AIC}_{model} - \mathrm{AIC}_{\La\mathrm{CDM}}$ for each distinct model and similarly for $\Delta$BIC.}
\label{theor}
\end{table}

\begin{figure*}[hbt!]
\centering
\begin{subfigure}
    \centering 
    \includegraphics[width=0.49\textwidth]{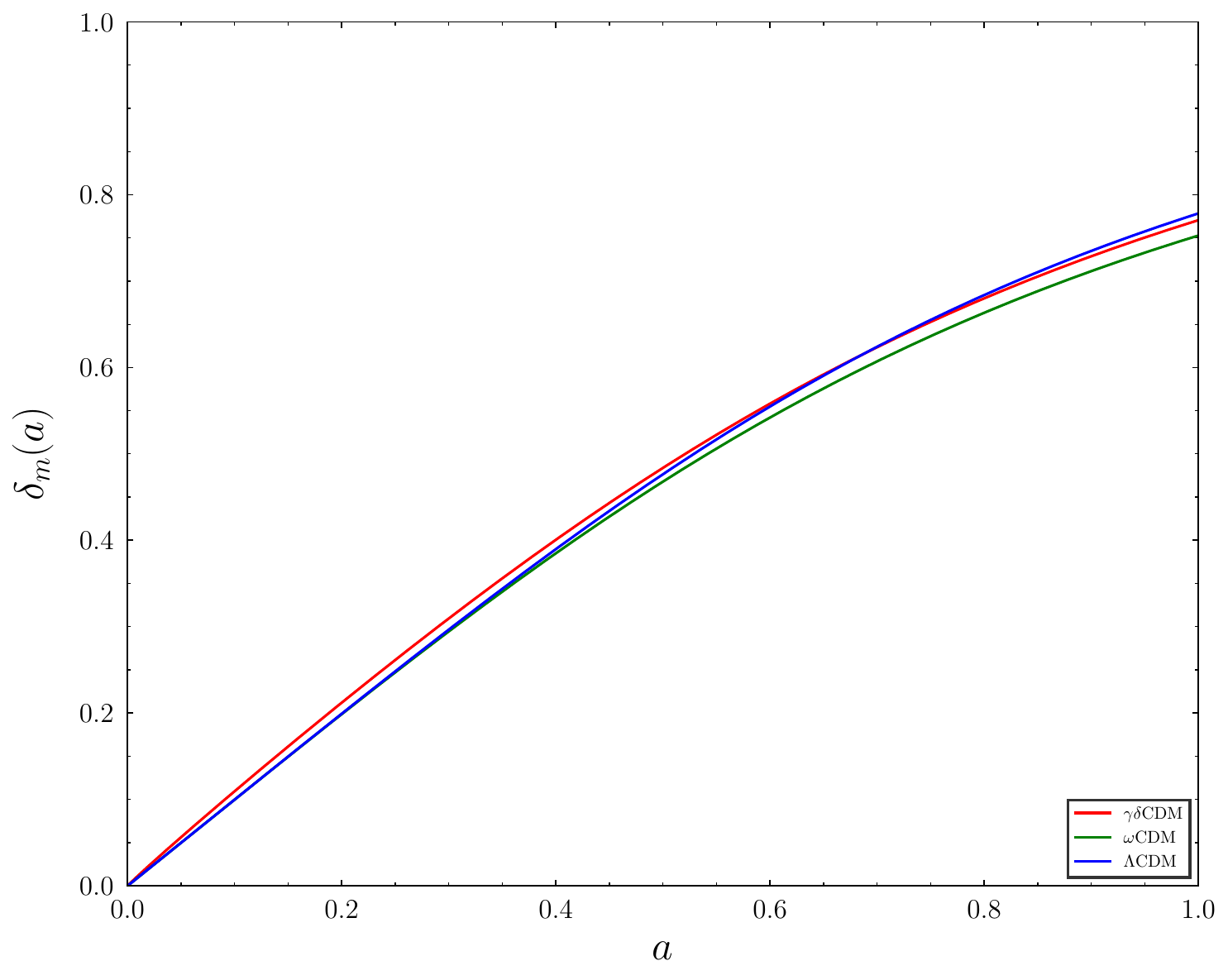}
    \label{fig:sub1a}
\end{subfigure}
\begin{subfigure}
    \centering 
    \includegraphics[width=0.49\textwidth]{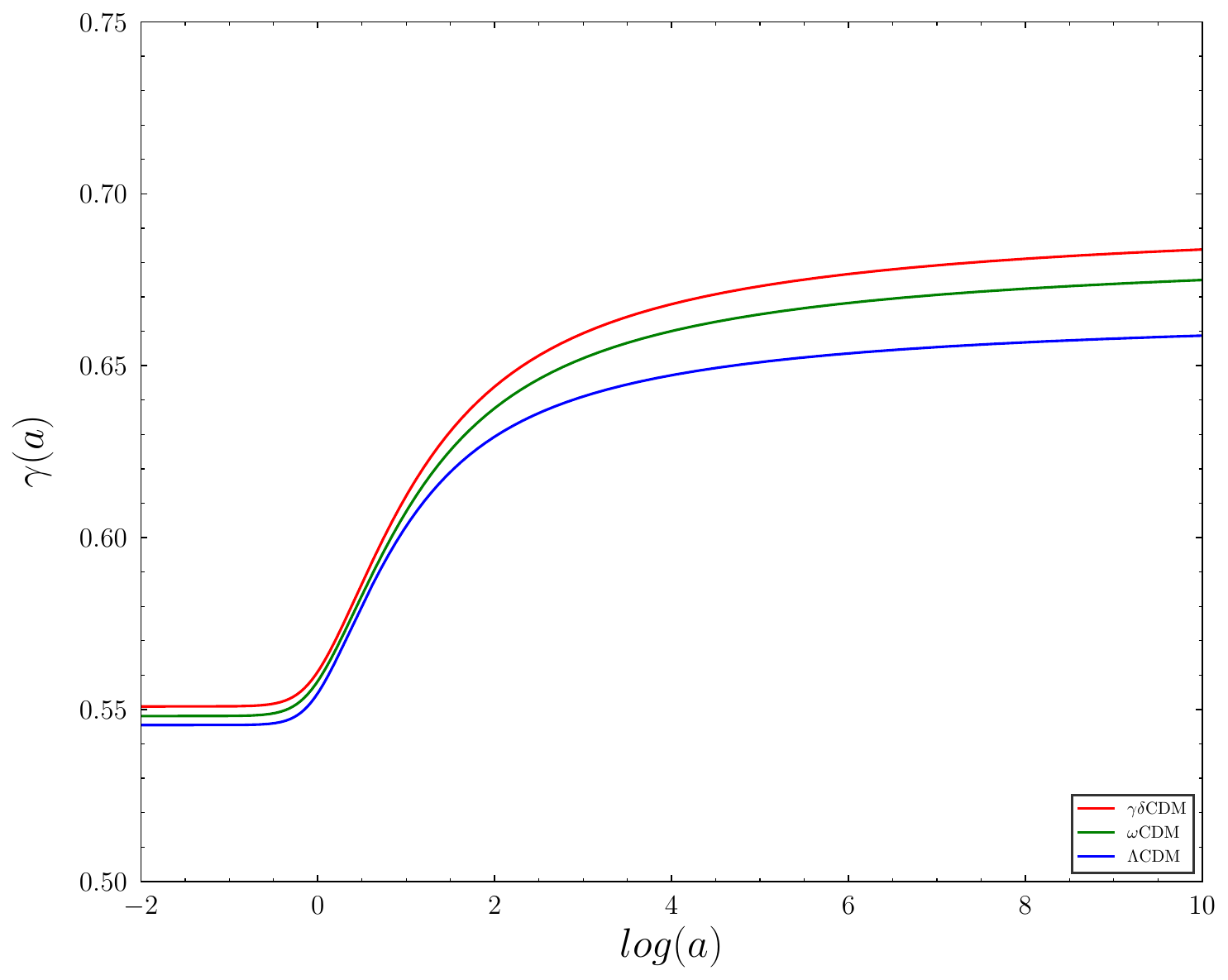}
    \label{fig:sub1b}
\end{subfigure}
\caption{Left panel: The density contrast, $\delta_m(a)$, is plotted versus scale factor, $a$. Equation (\ref{dmgs}) is used for the \gdcdm\ model and the growing solution given in \cite{Silveira:1994yq,Lee:2009gb,Nesseris:2017vor,Velasquez-Toribio:2020xyf} is used for the \wcdm\ and the \lcdm\ models. Right panel: The growth index undergoes a sudden increase near the present time. As in the \lcdm\ and the \wcdm\ models \cite{Linder:2018pth} its value remains constant in the matter dominated era and then asymptotically approaches to new numerical value in the far future. In both panels, we use values of the cosmological parameters obtained from the analyses with $f(z)$+DESI2+DESY5 dataset.}
\label{fzasym}
\end{figure*}

In all analyses, the chi-squared function for the \gdcdm\ and the \wcdm\ models have lower values compared to the chi-squared function of the \lcdm\ model. In that respect, the \gdcdm\ and the \wcdm\ models are preferred over the \lcdm\ model. However, these models have different number of free parameters. In order to take into account this disparity, one utilizes the model comparison criteria that penalize the existence of extra free parameters. The Akaike information criterion (AIC) \cite{Akaike:1974vps} and the Bayesian information criterion (BIC) \cite{Schwarz:1978vps} are the most widely used information criteria. They are defined by AIC $= 2k - 2\ln\mathcal{L}$ and BIC $= k\ln N - 2\ln\mathcal{L}$. Here $\mathcal{L}$ is the maximum value of the likelihood function (\ref{like}), $k$ is the number of model parameters, and $N$ is the number of data points. The penalty is larger in BIC than in AIC \cite{Stoica:2004vps} for large dataset sizes. The \lcdm\ model has one free parameter ($\Omega_{m0}$), the \wcdm\ model has one more ($\omega$), and the \gdcdm\ model has two more free parameters ($\omega$ and $\delta$). The chi-squared function for the $f(z)$ and the $f(z)$+DESI2 datasets have very close values for the three cosmological models. For these datasets, AIC and BIC values show preference for the \lcdm\ model. The $f(z)$+DESI2+DESY5 dataset combination have much larger number of data points, and therefore constrain the cosmological parameters much better. In the case of analysis with this large dataset we observe that AIC values show preference for the \gdcdm\ and the \wcdm\ models over the \lcdm\ model. However, since BIC penalizes the model with a larger number of parameters more, according to the BIC values, the \gdcdm\ model is not preferred over the other models. 

The density contrast, $\delta_m(a)$, is plotted versus the scale factor, $a$, in the left panel of Figure \ref{fzasym}. We use equation (\ref{dmgs}) for the \gdcdm\ model and the growing solution given in \cite{Silveira:1994yq,Lee:2009gb,Nesseris:2017vor,Velasquez-Toribio:2020xyf} for the \wcdm\ and the \lcdm\ models. We use values of the cosmological parameters obtained from the analyses with the $f(z)$+DESI2+DESY5 dataset in this plot. The similar behavior of the density contrasts in each model demonstrates that the growth of structure in the \gdcdm\ model follows a pattern similar to that of the \lcdm\ and the \wcdm\ models. This can also be seen as a test of the density contrast formula (\ref{dmgs}). As $a \ra 0$, the density contrast behaves as $\delta_m(a) \approx a^{(1-\delta)}$ in the \gdcdm\ model.

In the right panel of Figure \ref{fzasym}, the growth index $\gamma(a)$ is plotted with respect to $\log(a)$. In the asymptotic past, in the matter dominated era, the growth indices for the \gdcdm , the \lcdm\ and the \wcdm\ models approach constant values, they then undergo evolution about the present epoch, and later in the asymptotic future they approach higher constant values. The constant asymptotic values of the growth indices far ago ($a=10^{-3}$) are 0.545, 0.548 and 0.551 for the \lcdm , the \wcdm\ and the \gdcdm\ models, respectively. Whereas in the distant future ($a=10^{10}$), the constant asymptotic values of the growth indices are 0.656, 0.675 and 0.684 for the \lcdm , the \wcdm\ and the \gdcdm\ models, respectively. The behavior of the growth index versus $\log(a)$ could be significantly different in the $f(R)$ gravity models as shown in \cite{Linder:2018pth}. The fact that the growth index, $\gamma(a)$, of the \gdcdm\ model (\ref{hgri}) behaves similarly to the growth indices of the \lcdm\ and the \wcdm\ models distinguishes the \gdcdm\ model from the other $f(R)$ gravity models. We repeat here the crucial fact that the \gdcdm\ model is an exact solution of the modified Friedmann equations of $f(R)$ gravity. Most crucially, it does not source the dark energy contribution from some effective energy-momentum tensor constructed out of curvature terms.

\subsection{Analyses with free parameter $\gamma$ \label{numer}}

In the analyses, the results of which we present in this section, we do not evaluate the growth index from its theoretical formulae (\ref{gri}) and (\ref{hgri}), but leave it as a free parameter, to determine whether there is tension between the growth index determined by the data and the theoretical prediction of it by the specific cosmological model. Such analyses were performed in \cite{Nguyen:2023fip} and it was concluded that the cosmological datasets they used prefer much higher growth index compared to the prediction of the \lcdm\ model. 

We fit the cosmological models, together with free parameter $\ga$ for the growth index, to the combinations of the $f(z)$ growth rate data \cite{Avila:2022xad}, the DESI2 BAO data \cite{DESI:2025zgx}, and the DESY5 SNe Ia data \cite{DES:2024jxu,DES:2024hip,DES:2024upw,desy5_github}. 
Table \ref{num} displays the posterior values of each model's cosmological parameters and the growth index.  The difference between the chi-squared function of the particular model and the \lcdm\ one is also included in the same table.  Since there are not many data points in the case of analyses using only the $f(z)$ growth rate data, the posterior $1\si$ confidence areas are rather large.  Nevertheless, the models are significantly better constrained when the $f(z)$ growth rate data is combined with the DESI2 BAO data and the DESY5 SNe Ia data, with smaller $1\si$ confidence areas.

\begin{table}[hbt!]
\centering
\def\arraystretch{1.5}
\resizebox{\textwidth}{!}{%
\begin{tabular}{|l|c|c|c|c|c|c|c|c|}
\hline 
\multicolumn{1}{|l|}{} & \multicolumn{1}{|l|}{} & \multicolumn{4}{c|}{\textbf{Parameters}} & \multicolumn{3}{c|}{\textbf{Criteria}} \\
\hline 
\hline 
\textbf{Dataset} & \textbf{Models} & $\ga (a=1)$ & $\Omega_{m0}$ & $\omega$ & $\delta$ 
& $\Delta\chi^2$ & $\Delta$AIC & $\Delta$BIC \\
\hline 
\hline
 & \gdcdm & $0.520_{-0.082}^{+0.101}$ & $0.335_{-0.060}^{+0.044}$ 
& $-0.737_{-0.135}^{+0.189}$ & $0.077_{-0.054}^{+0.078}$ & $-0.08$ & 3.92 & 4.72   \\
f(z) & \wcdm & $0.549_{-0.090}^{+0.094}$ & $0.298_{-0.060}^{+0.062}$  & $-0.820_{-0.288}^{+0.273}$ 
& | & $-0.08$ & 1.92 & 2.32 \\
 & \lcdm & $0.561_{-0.091}^{+0.091}$ & $0.278_{-0.050}^{+0.054}$ & $-1$ & | & $0$ & $0$ & $0$   \\
\hline
 & \gdcdm & $0.544_{-0.069}^{+0.071}$ & $0.300_{-0.008}^{+0.008}$ 
& $-0.896_{-0.028}^{+0.029}$ & $0.047_{-0.032}^{+0.041}$ & $-1.38$ & 2.62 & 4.98 \\
f(z)+DESI2 & \wcdm & $0.576_{-0.063}^{+0.063}$ & $0.297_{-0.008}^{+0.008}$  & $-0.916_{-0.030}^{+0.029}$ 
& | & $-1.51$ & 0.49 & 1.67 \\
 & \lcdm & $0.596_{-0.062}^{+0.058}$ & $0.297_{-0.003}^{+0.003}$ & $-1$ & | & $0$ & $0$ & $0$ \\
\hline
\multirow{2}{*}{\multirow{2}{*}{\shortstack{\(f(z)\) + DESI2 \\ + DESY5}}} 
 & \gdcdm & $0.540_{-0.069}^{+0.072}$ & $0.299_{-0.008}^{+0.007}$ 
& $-0.890_{-0.027}^{+0.029}$ & $0.046_{-0.032}^{+0.043}$ & $-13.37$ & $-9.37$ & 1.68  \\
 & \wcdm & $0.569_{-0.062}^{+0.064}$  & $0.295_{-0.008}^{+0.008}$ & $-0.902_{-0.028}^{+0.027}$  & | & $-11.99$ 
 & $-9.99$ & $-4.47$ \\
 & \lcdm & $0.599_{-0.062}^{+0.058}$ & $0.300_{-0.003}^{+0.003}$ & $-1$ & | & $0$ & $0$ & $0$  \\
\hline
\hline
\end{tabular} 
}
\caption{The marginalized posterior values of the cosmological parameters and growth index, with 1$\sigma$ confidence regions, of each model obtained after the data analyses with three distinct datasets. $\Delta\chi^2$ compares each model's chi-squared function to the \lcdm's chi-squared function: $\Delta\chi^2 = \chi^2_{model} - \chi^2_{\La\mathrm{CDM}}$. Differences of information criterion values are calculated with $\Delta \mathrm{AIC} = \mathrm{AIC}_{model} - \mathrm{AIC}_{\La\mathrm{CDM}}$ for each distinct model and similarly for $\Delta$BIC.}
\label{num}
\end{table}

\begin{figure*}[hbt!]
\centering
\begin{subfigure}
    \centering 
    \includegraphics[width=0.45\textwidth]{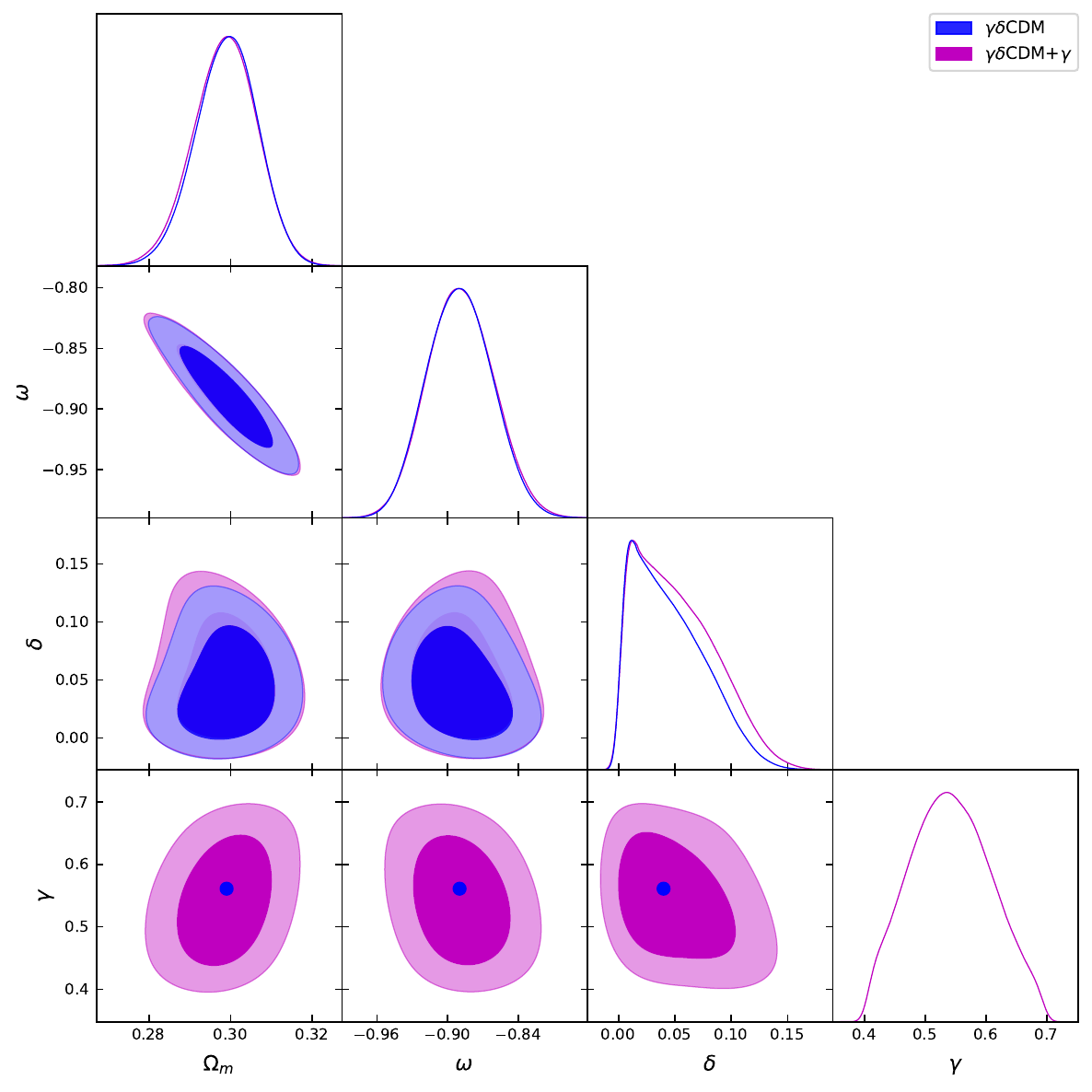}
    \label{fig:sub1a}
\end{subfigure}
\begin{subfigure}
    \centering 
    \includegraphics[width=0.49\textwidth]{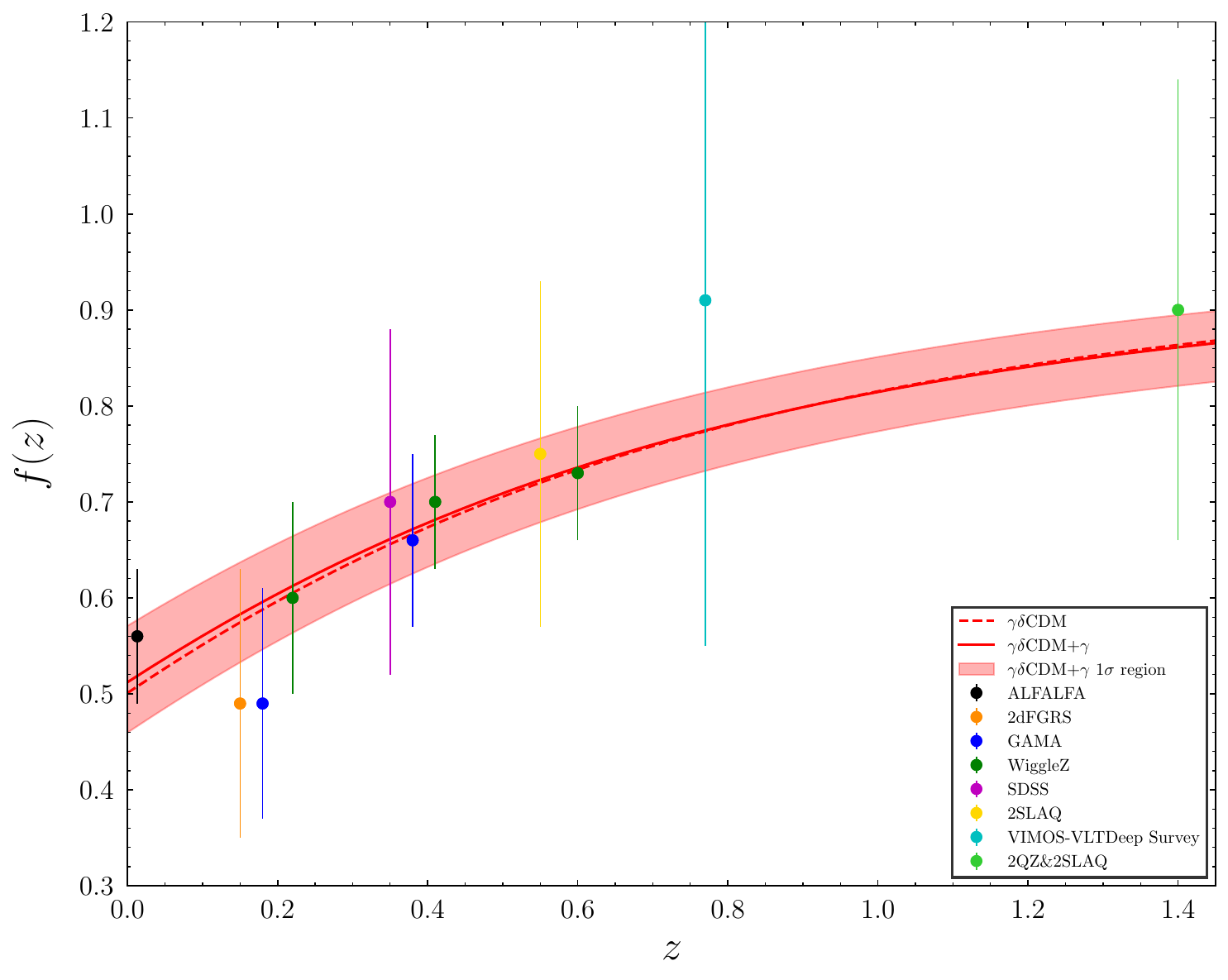}
    \label{fig:sub1b}
\end{subfigure}
\begin{subfigure}
    \centering 
    \quad \includegraphics[width=0.4\textwidth]{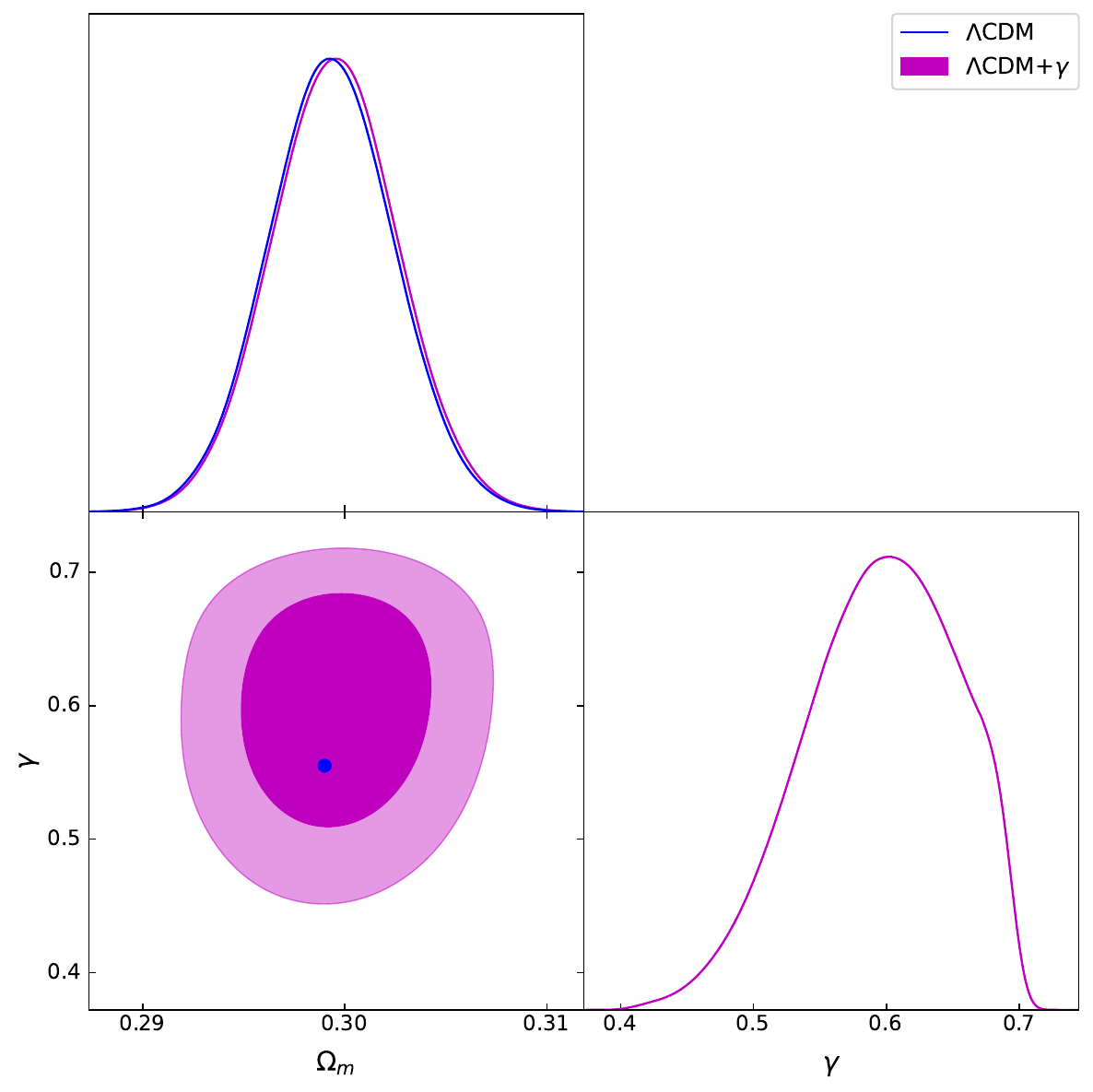}
    \label{fig:sub2a}
\end{subfigure}
\begin{subfigure}
    \centering 
    \qquad \includegraphics[width=0.49\textwidth]{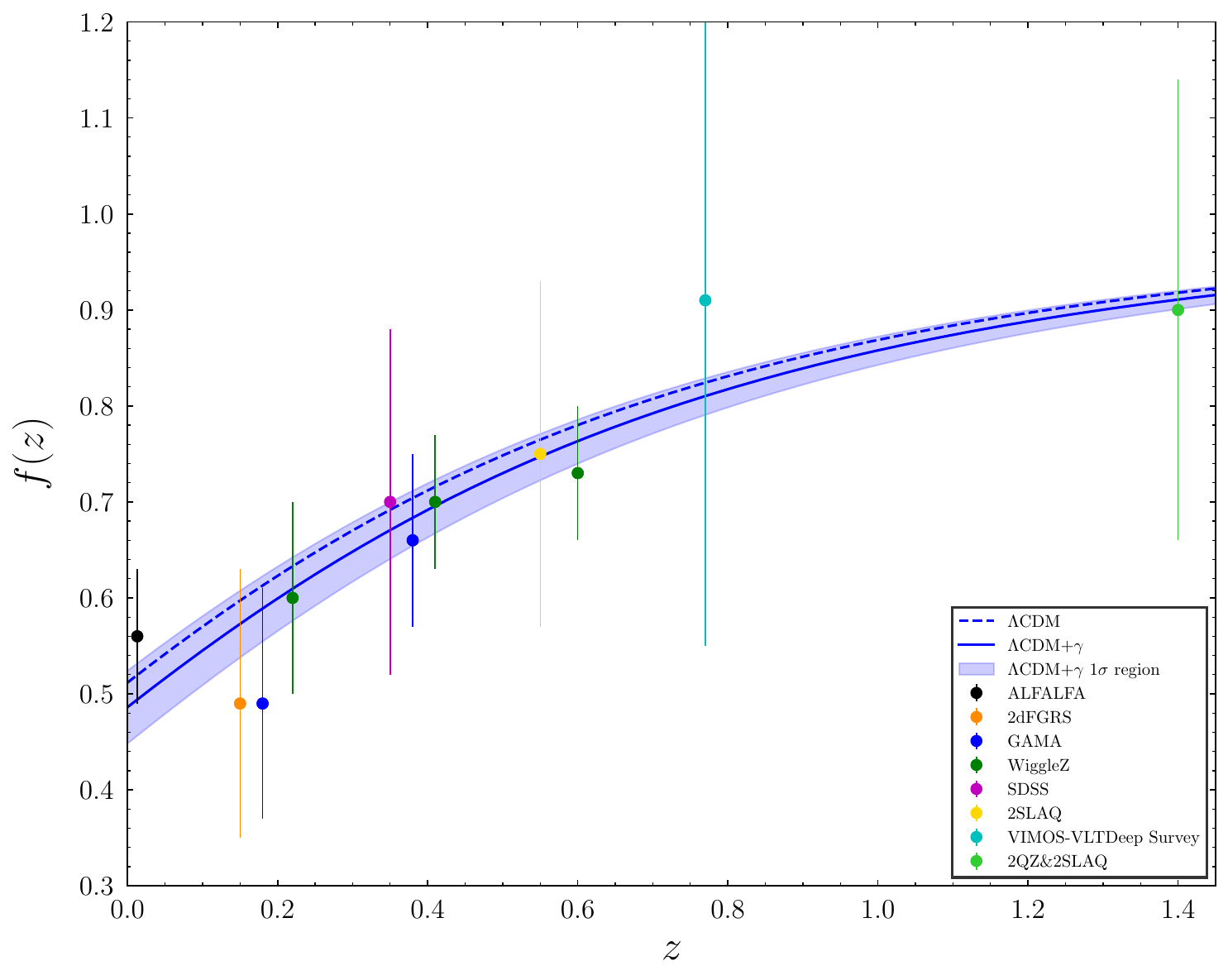}
    \label{fig:sub2b}
\end{subfigure}
\begin{subfigure}
    \centering 
    \quad \includegraphics[width=0.4\textwidth]{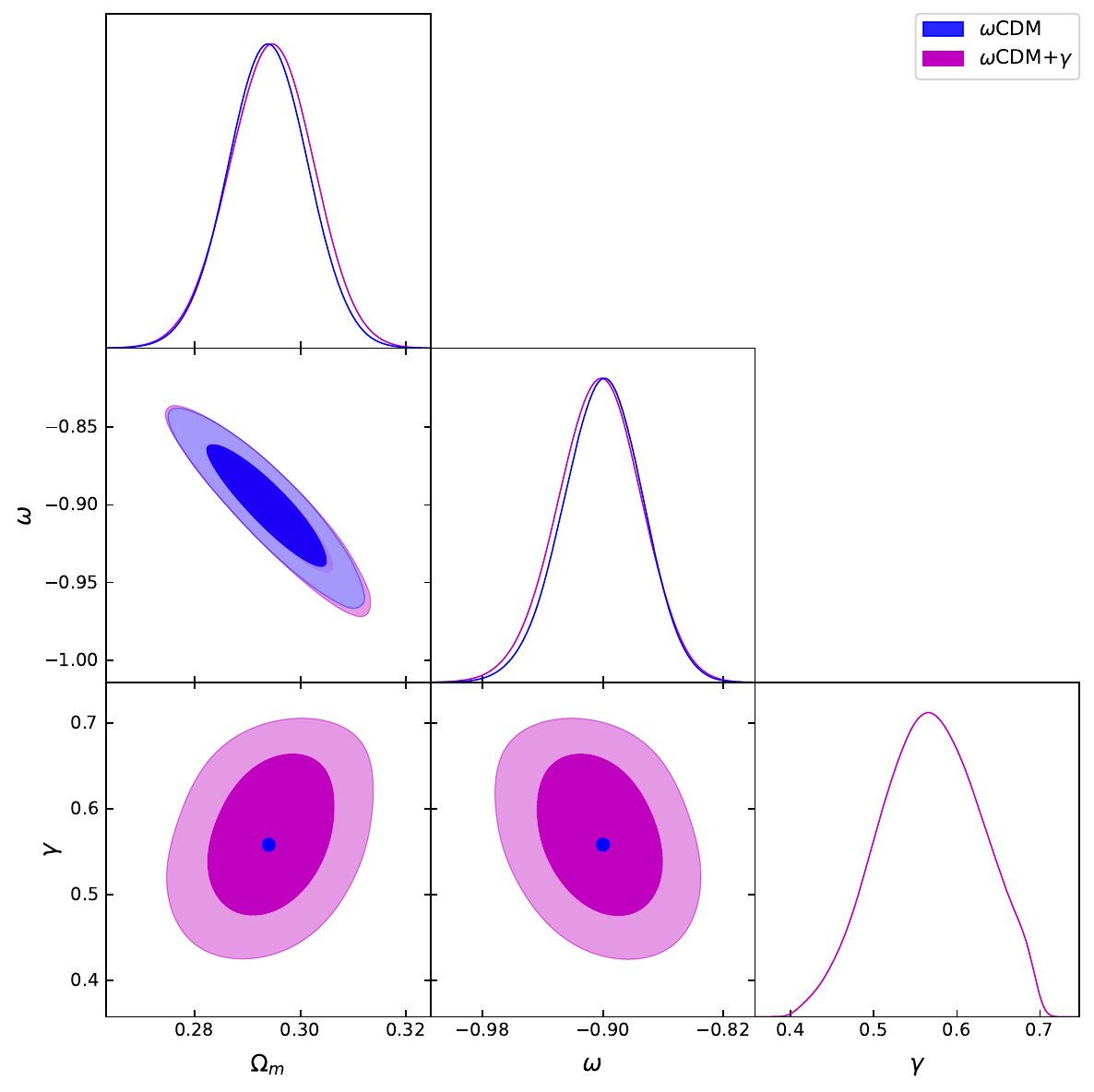}
    \label{fig:sub3a}
\end{subfigure}
\begin{subfigure}
    \centering 
    \qquad \includegraphics[width=0.49\textwidth]{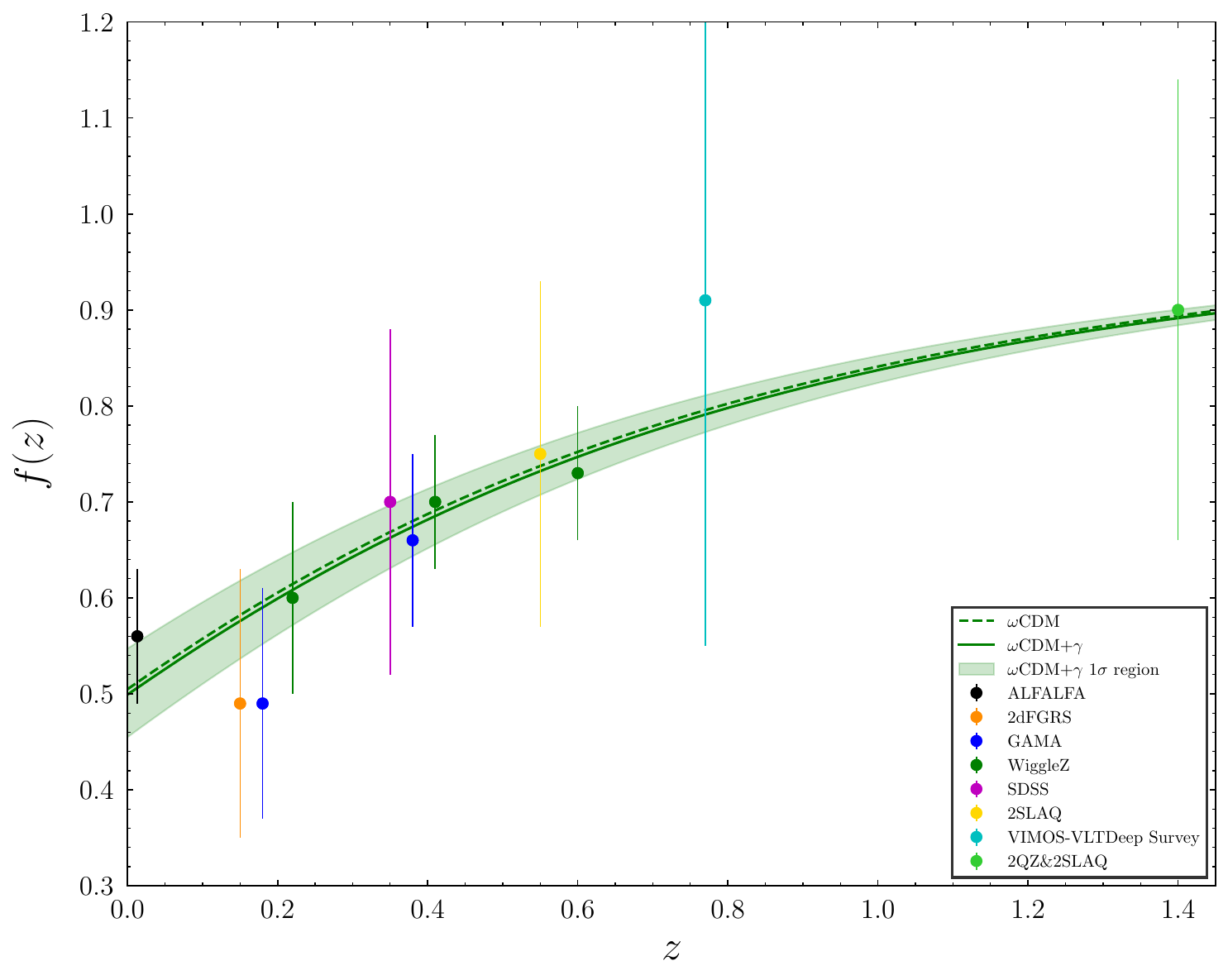}
    \label{fig:sub3b}
\end{subfigure}
\caption{Left panels: The contour plots for 2D joint posterior distributions, with 1$\sigma$ and 2$\sigma$ confidence regions, for only free cosmological parameters (blue) and also together with the free growth indices (purple) of the \gdcdm, the \lcdm\ and \wcdm\ models, for the full $f(z)$+DESI2+DESY5 dataset, together with 1D marginalized posterior distributions. Right panels: Theoretical (dashed curves) and the numerically determined (solid curves) growth rate functions plotted together with the $f(z)$ growth rate data (see Table \ref{fzdata}) for the \gdcdm, the \lcdm\ and \wcdm\ models. Shaded bands indicate $1\si$ posterior regions.}
\label{fzall}
\end{figure*}

As in the previous section, the difference of the chi-squared functions, $\Delta\chi^2 = \chi^2_{model} - \chi^2_{\La\mathrm{CDM}}$, indicates that the \gdcdm\ and the \wcdm\ models are preferred over the \lcdm\ model in all analyses. It is also worth noting that the inclusion of the DESY5 SNe Ia data enhances the preference of the \gdcdm\ model compared to the \wcdm\ model. In order to take into account the disparity in the number of free parameters of the distinct models, we again calculate the AIC and the BIC values. In this section, again, the chi-squared function for the $f(z)$ and the $f(z)$+DESI2 datasets have very close values for the three cosmological models. For these datasets, AIC and BIC values show preference for the \lcdm\ model. The $f(z)$+DESI2+DESY5 dataset combination have much larger number of data points, and therefore constrain the cosmological parameters much better. In the case of analysis with this large dataset we observe that AIC values show preference for the \gdcdm\ and the \wcdm\ models over the \lcdm\ model. However, since BIC penalizes the model with a larger number of parameters more, according to the BIC values, the \gdcdm\ model is not preferred over the other models.

Figure \ref{fzall} includes on its left panels from top to bottom the contour plots for 2D joint posterior distributions, with 1$\sigma$ and 2$\sigma$ confidence regions, for only free cosmological parameters (blue colored) and also together with the free growth indices (purple colored) of the \gdcdm, the \lcdm\ and \wcdm\ models, for the full $f(z)$+DESI2+DESY5 dataset, together with 1D marginalized posterior distributions. These contour plots show that theoretical formulae for the growth index and the cosmological data are in excellent agreement. To perceive this agreement visually, on the right panels of Figure \ref{fzall} we plot the theoretical (dashed curves) and the numerically determined (solid curves) growth rate functions together with the $f(z)$ growth rate data (see Table \ref{fzdata}) for the \gdcdm, the \lcdm\ and \wcdm\ models from top to bottom. Shaded bands in these plots indicate $1\si$ posterior regions. For all the models, the theoretically predicted values of the growth indices are in the $1\si$ region of the numerically determined ones. Furthermore, in the cases of the \gdcdm\ and the \wcdm\ models the numerically and the theoretically determined curves are extremely close to each other. Thus, data strongly support the theoretical formulae for the growth rate in these models. The larger $1\sigma$ region in the graph of the \gdcdm\ model's $f(z)$ function versus redshift is due to the existence of the growth rate calibration parameter, $f_\infty = (1-\delta)$ (\ref{fgex}).

\subsection{Comparison with other models \label{models}} 

As argued in \cite{Linder:2007hg} and \cite{Polarski:2007rr}, distinct dark energy models with general relativistic description of gravity have growth indices not too distinct from the \lcdm's one, $\gamma \sim 0.55$. In contrast, in cosmological models based on alternative or modified gravity theories, the growth index can have quite distinct values. This prediction has been supported by many works in the later literature. In \cite{Linder:2007hg} the growth index in Dvali-Gobadadze-Porrati (DGP) gravity \cite{Dvali:2000hr,Deffayet:2001pu} is calculated from the derived fitting formula to be $\gamma = 11/16 \approx 0.688$. However, when the growth index is treated as a free parameter of the model and then the model with this extra parameter is confronted with data, both the \lcdm\ model's and the DGP gravity's growth indices are found to be very different compared to their expected theoretical values: In \cite{Gong:2008fh} it is found that the observational constraints force the growth index in the \lcdm\ model to be $\gamma = 0.64^{+0.17}_{-0.15}$ and the growth index in the DGP gravity model to be $\gamma = 0.55^{+0.14}_{-0.13}$. When the \lcdm\ model is fitted to more recent datasets, the growth index persisted to be substantially larger than its theoretically expected value. In \cite{Nguyen:2023fip}, \lcdm\ growth index, set as a free parameter in the analyses, is determined to be $\gamma = 0.598^{+0.031}_{-0.031}$ in the analysis with $f\si_8$+DESY1+BAO data combination, and $\gamma = 0.633^{+0.025}_{-0.024}$ in the analysis with PL18+$f\si_8$+DESY1+BAO data combination. Lately, Giar\`e at al. \cite{Giare:2025ath} analyzed the \lcdm\ model with the growth index and the sum of the neutrino masses ($\sum m_\nu$) as free parameters. They found that the growth index needs to be as large as $\gamma = 0.742 \pm 0.069$ for Plik+DESI2+Pan+ data combination when enforcing a NO prior on $\sum m_\nu$, and $\gamma = 0.707 \pm 0.075$ for the same data combination when not assuming a NO prior on $\sum m_\nu$ (without including Cepheid host distances in the Pantheon+ SNe Ia dataset \cite{Giare:2025ath}). In this work, we determine $\gamma = 0.599_{-0.062}^{+0.058}$ when \lcdm\ together with $\gamma$ treated as a free parameter. 
This value of the growth index is small compared to what is determined in \cite{Nguyen:2023fip,Giare:2025ath} in the analyses that include Planck CMB data. It thus seems that including the Planck CMB data forces the growth index to have higher posterior values. As explained in the Introduction and Section \ref{data}, we deliberately avoided using the Planck CMB data so that we can isolate the effect of the Hubble tension from the growth of structure. The value of the growth index in the \gdcdm\ model obtained both from the theoretical formula (\ref{fQ}) and from the data analysis is close to the theoretical value of the growth index in the \lcdm\ or the \wcdm\ models.

\begin{figure*}[hbt!]
\centering
\begin{subfigure}
    \centering 
    \includegraphics[width=0.49\textwidth]{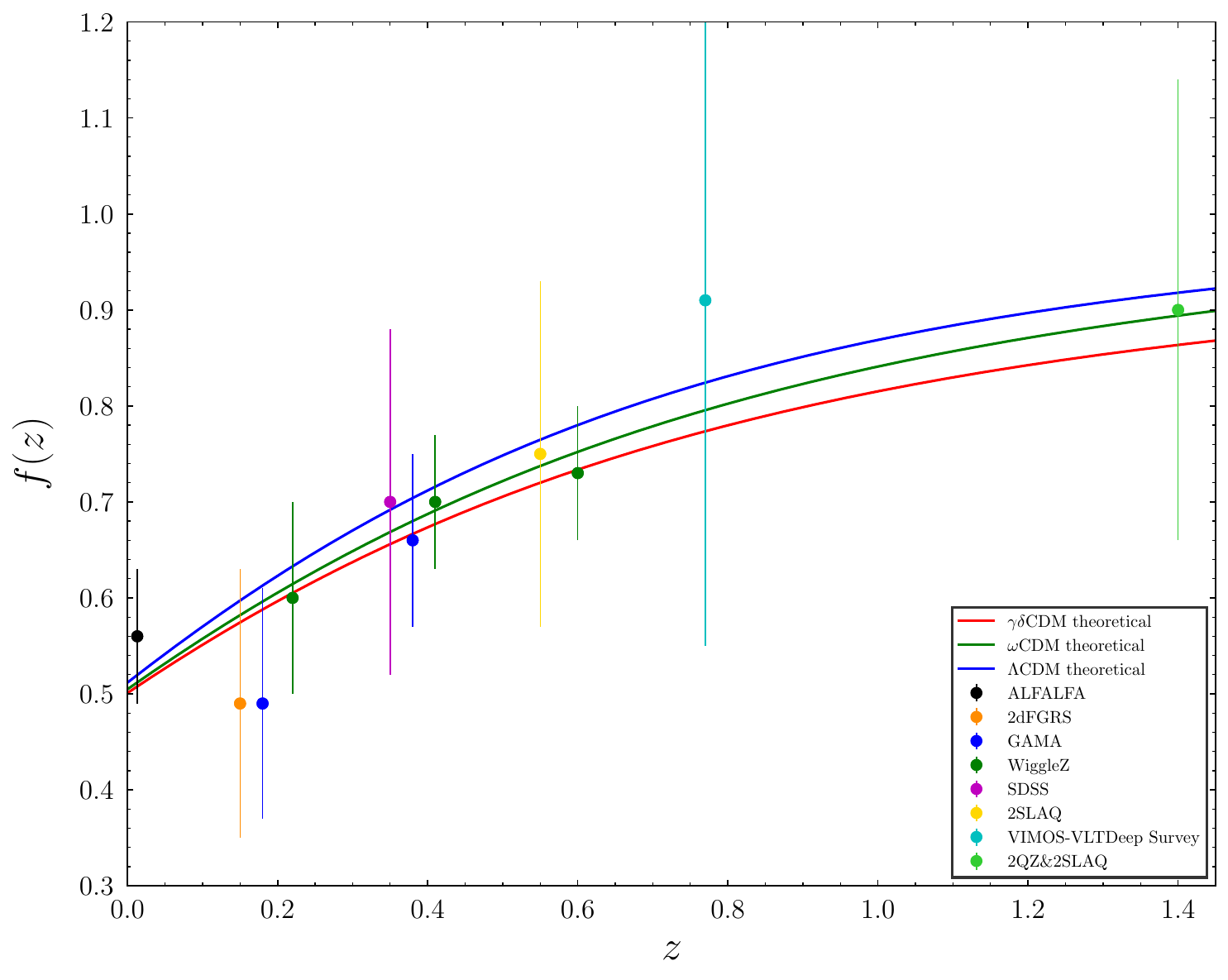}
    \label{fig:sub1a}
\end{subfigure}
\begin{subfigure}
    \centering 
    \includegraphics[width=0.49\textwidth]{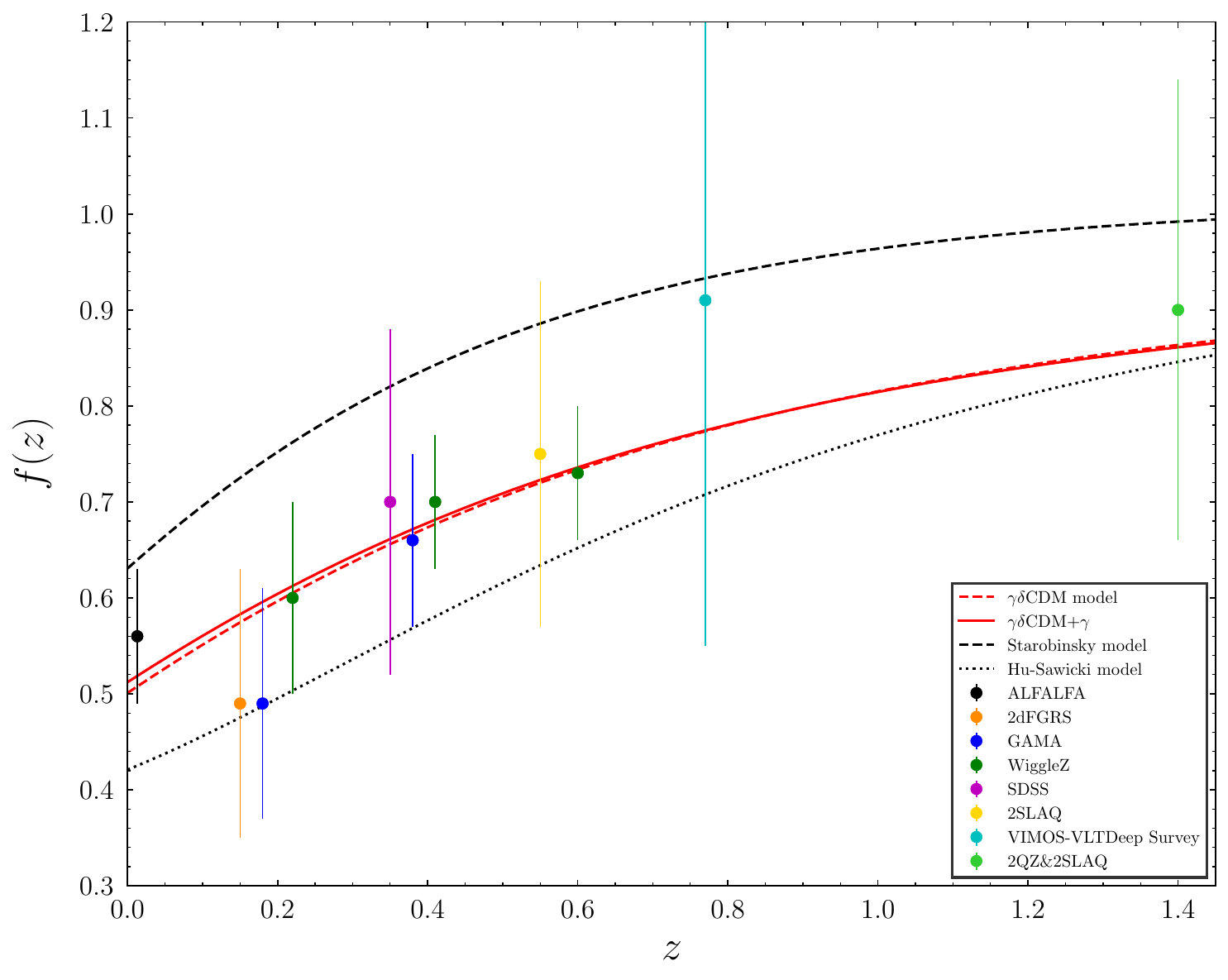}
    \label{fig:sub1b}
\end{subfigure}
\caption{Comparison of theoretical growth rate $f(z)$ for various models. On the left panel the \gdcdm\ model is compared with the \lcdm\ and the \wcdm\ models. On the right panel the \gdcdm\ model is compared with the $f(R)$ gravity models. It is visually obvious that the growth rate in the \gdcdm\ model behaves similarly to the \lcdm\ and the \wcdm\ models, and distinctly to the $f(R)$ gravity models, even though \gdcdm\ model is based on $f(R)$ gravity.}
\label{fzcom}
\end{figure*}

As pointed out in \cite{Linder:2007hg} the growth index in alternative gravity theories is scale dependent and the difference from the growth index of the \lcdm\ model is given in terms of the parameterized post-Newtonian (PPN) parameter $\ga_{PPN}$ \cite{Linder:2007hg,Will:2005va,Anton:2025zer}. It is worth mentioning the growth index values in two distinct $f(R)$ gravity models to observe this scale dependence: In \cite{Gannouji:2008wt} the growth index is determined for the Starobinsky model \cite{Starobinsky:2007hu} and is found to have the value $\ga (z) = 0.399  -0.246 z$ for $\Omega_{m0} = 0.315$. In \cite{Tsujikawa:2009ku} it is argued that the value of the growth index is found to be $\ga \sim 0.4$ for many viable $f(R)$ gravity models. In contrast, in the Hu-Sawicki model \cite{Hu:2007nk,Basilakos:2013nfa} the growth index is found \cite{Basilakos:2017rgc} to be $\ga (a) = 0.753 + 0.690 (1-a)$, which is calculated at a different scale. 
Two more f(R) gravity models are analyzed in \cite{Bamba:2012qi} with $\ga \sim 0.46$ and $\ga \sim 0.51$. However, in this work, the growth rate approaches 1 at high redshift, unlike the other $f(R)$ models, for which $f > 1$ in the same limit.
The theoretical value of the growth index in the $f(R)$ based \gdcdm\ model, $\ga (a=1) = 0.561$, is very different from the value of the growth index in other $f(R)$ gravity based cosmological models. As shown in the right panel of Figure \ref{fzcom}, the growth rate (\ref{fgOa}) evaluated with the specified value of the growth index fits the growth rate data much better compared to the other $f(R)$ gravity models.

However, the difference with the other $f(R)$ gravity models is not just the different value of the growth index. In the \gdcdm\ model, we need to utilize the extended gravitational growth framework \cite{Linder:2009kq,Linder:2013dga}, which is forced by the non-standard high redshift limit of the density contrast, $\delta_m \propto a^{(1-\delta)}$. Thus, the growth rate has to be defined by including the growth rate calibration parameter, $f_\infty$, as $f(a) = f_\infty \hat{\Omega}_m^\gamma$ (\ref{finf}). We find in Section \ref{index} that the growth rate at high redshift is less than what is found in models based on general relativity: $f_\infty = (1-\delta) < 1$ (\ref{fQ}) in the \gdcdm\ model.

In contrast, the growth rate in some $f(R)$ gravity models is larger than the general relativistic counterparts \cite{Linder:2009kq,Linder:2013dga,Cai:2013toa}. In those models, $f > 1$ at high redshift is due to deeper gravitational wells caused by the extra attractive force of the scalar field. The linear ISW effect is intimately related to the growth rate of structure because both depend on the evolution of the gravitational wells \cite{Cai:2013toa,Cooray:2003hd,Fang:2008kc,Beck:2018owr}. Effectively, the higher growth rate slows the decay of the potential wells more than general relativistic models, and causes the ISW effect to be smaller on large scales \cite{Cai:2013toa}. This is true for both the early and late ISW effects, although due to increased dominance of cosmic expansion, the late ISW effect is close to the general relativistic one, but still smaller. Thus, usual $f(R)$ models cannot account for the observed high magnitude of the ISW signal \cite{Cai:2013toa,Granett:2008ju}.

In the \gdcdm\ model, $f \ra f_\infty = (1-\delta) < 1$ at high redshift indicates that the ISW effect will be completely different. Here, a smaller growth rate at all redshifts means that growth is suppressed compared to general relativistic and the other $f(R)$ models. In the early Universe, suppression of growth points to dominance of cosmic expansion over the growth of structure. This is similar to what happens in the EDE model. The Universe expands more rapidly at high redshift and this diminishes the growth. The lower growth rate and effect of cosmic expansion would speed up the decay of the potential wells and cause the ISW effect to be higher on large scales. Thus, we expect that both the early and late ISW effects to be enhanced in the \gdcdm\ model. This enhancement will be less in the late Universe as the growth rate behaves similarly to the \lcdm\ and the \wcdm\ models at low redshift (see the left panel of Figure \ref{fzcom}). This conclusion is more in line with the observed high magnitude of the ISW signal \cite{Granett:2008ju}. 

\subsection{$\sigma_8$ and $S_8$ parameters \label{sig8}} 

An important question is whether the \gdcdm\ model might be of relevance to $S_8$ tension. We plan to analyze this question with both low- and high-redshift data in a future publication. The scope of the present paper is to determine the density contrast and the growth rate functions, as well as the growth index in the \gdcdm\ model. These were accomplished in the previous sections. As a preliminary to the aforementioned future work, we now want to assess the potential of the \gdcdm\ model in resolving the $S_8$ tension. For that purpose, we fit the \gdcdm\ model to the $\si_8 (z)$ \cite{Piccirilli:2024xgo,Oliveira:2025huk} and the $f\si_8 (z)$ \cite{Skara:2019usd,Kazantzidis:2018rnb} datasets. Since these are datasets with few data points, to increase statistical significance we also add the DESI2 BAO data \cite{DESI:2025zgx} and the DESY5 SNe Ia data \cite{DES:2024jxu,DES:2024hip,DES:2024upw,desy5_github} to the analyses.
For the purpose of model comparison, we further fit the \lcdm\ model with the free $\Omega_{m0}$ parameter and separately with the predetermined value of $\Omega_{m0} = 0.315$ from \cite{Planck:2018vyg}. We name the last analysis ``Planck'' to distinguish from \lcdm\ analysis with free $\Omega_{m0}$ parameter.

\begin{table}[hbt!]
\centering
\renewcommand{\arraystretch}{1.5}
\resizebox{\textwidth}{!}{%
\begin{tabular}{|c|l|c|c|c|c|c|c|c|c|c|c|}
\hline
\hline 
\multicolumn{1}{|l|}{} & \multicolumn{1}{|l|}{} & \multicolumn{5}{c|}{\textbf{Parameters}} & \multicolumn{3}{c|}{\textbf{Criteria}} \\
\hline 
\textbf{Dataset}  & \textbf{Models} & $\Omega_{m0}$ & $\omega$ & $\delta$ & $\sigma_8$ & $S_8$ 
& $\Delta\chi^2$ & $\Delta$AIC & $\Delta$BIC \\
\hline
\hline
\multirow{2}{*}{\multirow{2}{*}{\shortstack{\(\sigma_8\) + DESI2 \\ + DESY5}}} 
& $\gamma\delta$CDM & $0.298_{-0.008}^{+0.008}$ & $-0.885_{-0.028}^{+0.028}$ & $0.047_{-0.033}^{+0.044}$ & $0.776_{-0.014}^{+0.014}$ & $0.773_{-0.014}^{+0.014}$ & $-13.95$ & $-9.95$ & 1.10 \\
& $\Lambda$CDM  & $0.299_{-0.003}^{+0.003}$ & - & - & $0.796_{-0.012}^{+0.012}$ & $0.795_{-0.012}^{+0.012}$ & 0 & 0 & 0 \\
& Planck  & $0.315$ & - & - & $0.802_{-0.011}^{+0.012}$ & $0.822_{-0.011}^{+0.012}$ & $23.86$ & 21.86 & 16.33 \\
\hline
\multirow{2}{*}{\multirow{2}{*}{\shortstack{\(f\sigma_8\) + DESI2 \\ + DESY5}}}
& $\gamma\delta$CDM &  $0.298_{-0.008}^{+0.008}$ & $-0.888_{-0.028}^{+0.028}$ & $0.046_{-0.033}^{+0.046}$ & $0.803_{-0.027}^{+0.029}$ & $0.800_{-0.027}^{+0.029}$ & $-12.55$ & $-8.55$ & 2.52 \\
& $\Lambda$CDM  & $0.299_{-0.003}^{+0.003}$ &  $-1$ & - & $0.758_{-0.020}^{+0.021}$ & $0.757_{-0.020}^{+0.021}$ 
& 0 & 0 & 0 \\
& Planck & $0.315$ & $-1$ & - & $0.752_{-0.020}^{+0.020}$ & $0.771_{-0.020}^{+0.020}$ & $23.72$ & 21.72 & 16.18 \\
\hline
\hline
\end{tabular}%
}
\caption{The marginalized posterior values of the $\si_8$ and the $S_8$ parameters, with 1$\sigma$ confidence regions, of each model obtained after the data analyses with two distinct datasets. $\Delta\chi^2$ compares each model's chi-squared function to the \lcdm's chi-squared function: $\Delta\chi^2 = \chi^2_{model} - \chi^2_{\La\mathrm{CDM}}$. Differences of information criterion values are calculated with $\Delta \mathrm{AIC} = \mathrm{AIC}_{model} - \mathrm{AIC}_{\La\mathrm{CDM}}$ for each distinct model and similarly for $\Delta$BIC.}
\label{tig8}
\end{table}

\begin{figure}[hbt!]
\centering
   \includegraphics[width=0.57\textwidth]{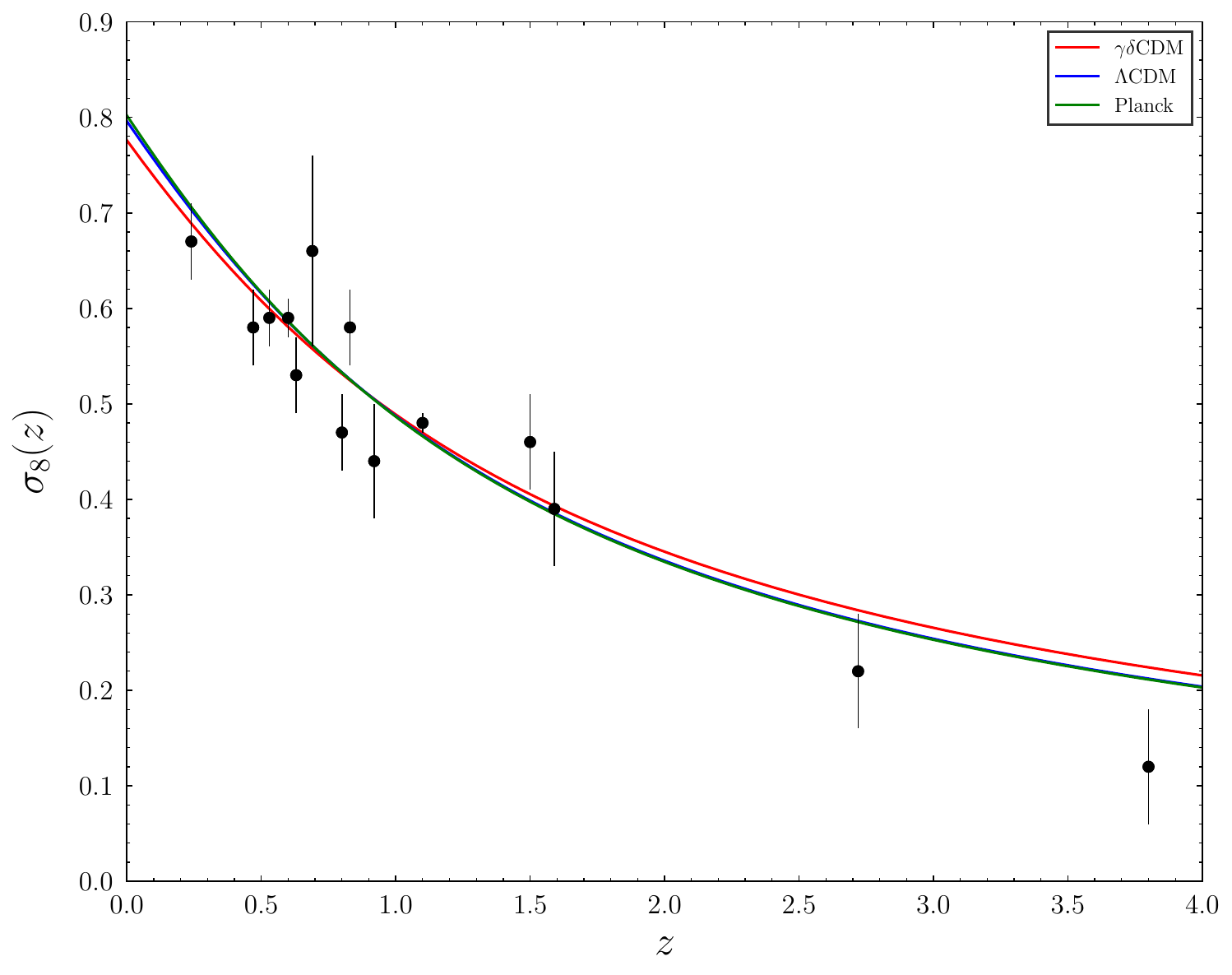}
\caption{The $\si_8(z)$ predictions of three cosmological models plotted together with the $\si_8$ data points of Table \ref{s8data}. Here the red curve corresponds to the \gdcdm\ model, blue curve to the \lcdm\ model with free $\Omega_{m0}$ parameter, and green curve to the \lcdm\ model with $\Omega_{m0}=0.315$, which is the value obtained by the Planck collaboration \cite{Planck:2018vyg}.}
\label{figs8}
\end{figure}

\begin{figure}[hbt!]
\centering
   \includegraphics[width=0.86\textwidth]{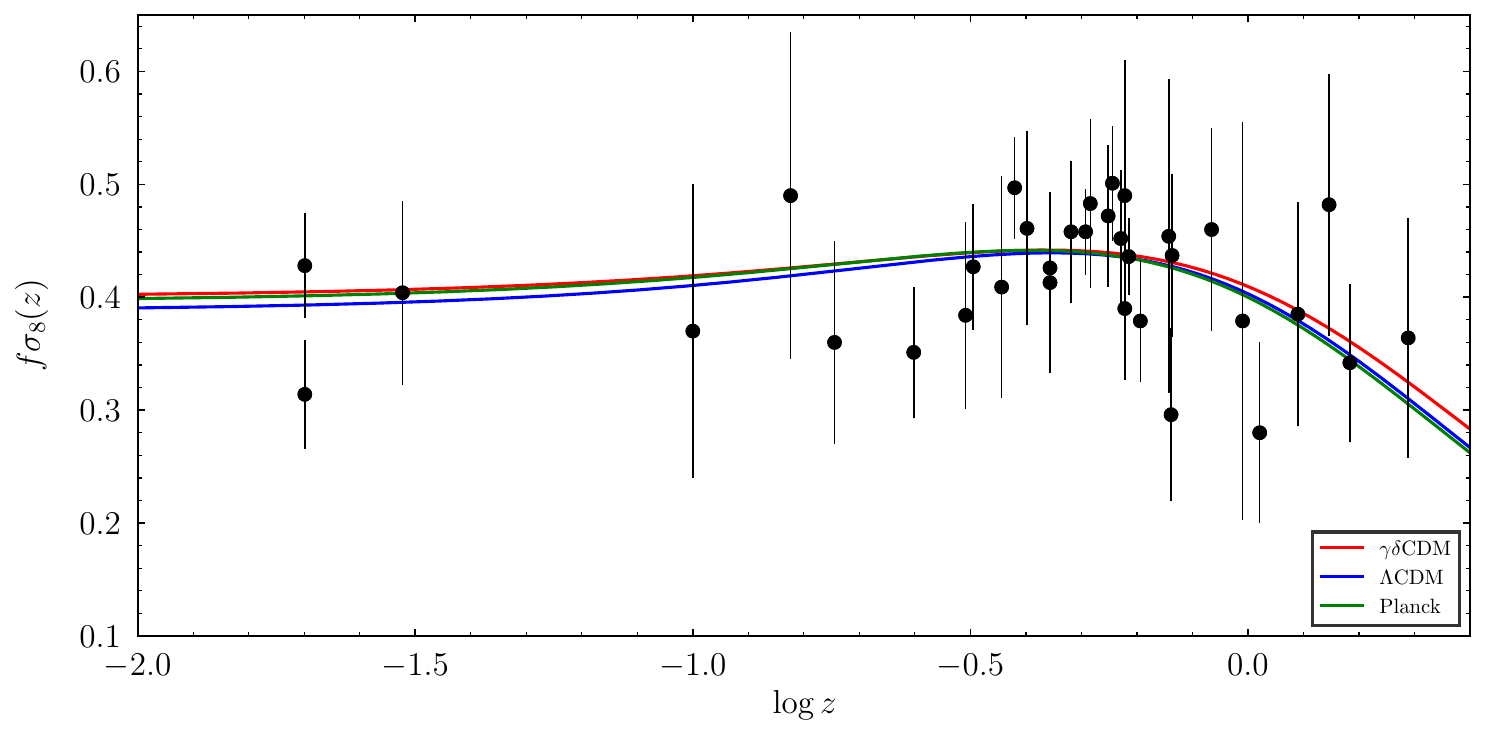}
\caption{The $f\si_8(z)$ predictions of three cosmological models plotted together with the $f\si_8$ data points of Table \ref{fsigma8}. Here the red curve corresponds to the \gdcdm\ model, blue curve to the \lcdm\ model with free $\Omega_{m0}$ parameter, and green curve to the \lcdm\ model with $\Omega_{m0}=0.315$, which is the value obtained by the Planck collaboration \cite{Planck:2018vyg}.}
\label{figfs8}
\end{figure}

For the \gdcdm\ model we utilize the theoretical formulae for the growth rate (\ref{fQ}) and the density contrast (\ref{dmgs}) to evaluate the functions $\sigma_8(z)$ and $f\sigma_8(z)$. For the \lcdm\ model we use the growth rate formula (\ref{fL}) and the density contrast (\ref{dmgs}) for $\delta = 0$ and $\omega = -1$ to evaluate the same functions. 
The posterior values of the cosmological parameters together with $\sigma_8$ and $S_8$ parameters for each model are presented in Table \ref{tig8}. We also present the difference in the information criteria values in the same table.

In all analyses, the difference in the values of the chi-squared function, $\Delta\chi^2 = \chi^2_{model} - \chi^2_{\La\mathrm{CDM}}$, indicates that the \gdcdm\ model is preferred over the \lcdm\ model. Since the \gdcdm\ model has two more free parameters compared to the \lcdm\ model, we also calculate the values of the Akaike (AIC) and Bayesian (BIC) information criteria for each model in each analysis. In the case of analyses with the $\sigma_8$+DESI2+DESY5 dataset combination the \gdcdm\ model is statistically preferred over the \lcdm\ model. The fixed value for $\Omega_{m0} = 0.315$ from the CMB data obtained by the Planck collaboration \cite{Planck:2018vyg} causes the fit of the \lcdm \ model to the late-time $\sigma_8(z)$ data to be worse.
The values of Bayesian information criterion show preference for the \lcdm\ model with the free $\Omega_{m0}$ parameter over the \gdcdm\ model. This is somewhat expected, since the penalty is higher in BIC than in AIC \cite{Stoica:2004vps} for large dataset sizes. 

In the case of analyses with $f\sigma_8 + DESI2 + DESY5$ dataset combination the Akaike information criterion again shows preference of the \gdcdm\ model over the \lcdm\ model. The fixed value for $\Omega_{m0} = 0.315$ from the CMB data obtained by Planck collaboration \cite{Planck:2018vyg} still causes the fit of the \lcdm\ model to the late-time $\sigma_8(z)$ data to be worse.
Here, Bayesian information criterion again prefers the \lcdm\ model with the free $\Omega_{m0}$ parameter over the \gdcdm\ model. 
The contrast between AIC and BIC results can again be attributed to the size of the dataset. 

As a visual representation we also plot the $\si_8(z)$ and the $f\si_8(z)$ functions for each model, using the posterior values of the cosmological parameters from Table \ref{tig8}, together with the $\si_8(z)$ and $f\si_8(z)$ data points, in figures \ref{figs8} and \ref{figfs8}, respectively. The closeness of the curves supports our conclusions in the previous section that the \gdcdm\ model behaves more similarly to the \lcdm\ model compared to other $f(R)$ models. Certainly, the analyzes in this section are not conclusive in deciding on the status of the $S_8$ tension in the \gdcdm\ model. However, they show that the \gdcdm\ model is indeed a competitive model for reproducing the $\si_8(z)$ and the $f\si_8(z)$ data from the observed inhomogeneity in the distribution of matter.

\section{Conclusions \label{conc}} 

Tensions in the current cosmological paradigm indicate that the current understanding of the Cosmos is not complete and that our cosmological model should be modified one way or the other. The resolutions of the cosmological tensions can either be found inside the current paradigm, or some radical change will bring us a better understanding and a better set of explanations. While dealing with the tensions it is imperative to distinguish the effects of different cosmological processes, each of which corresponds to one of the tensions. The Hubble constant is a crucial parameter when modeling the cosmological expansion. The related Hubble tension has not been resolved yet, albeit many efforts have been made \cite{CosmoVerse:2025txj}. Another important tension, the so called $S_8$ tension, is related to the growth of structure in the Universe. One needs to properly distinguish these two phenomena to correctly address each tension. As the value of the Hubble constant characterizes the Hubble tension, the value of the growth index at present time can be taken to characterize the $S_8$ tension \cite{Nguyen:2023fip}. 
As we stressed in the Introduction our aim in this article is not to investigate either the $H_0$ or the $S_8$ tensions. The scope of this work is to calculate the growth index in the \gdcdm\ model and to assess its relevance in alleviating the $S_8$ tension.

Different dark energy models with a general relativistic description of gravity have growth indices that are not all that different from the \lcdm's one, $\gamma \sim 0.555$, as stated in \cite{Linder:2007hg} and \cite{Polarski:2007rr}. On the other hand, the growth index might have quite different values in cosmological models that are based on alternative or modified gravity theories. According to \cite{Linder:2007hg}, the growth index in alternative gravity theories depends on the scale, and the difference from the growth index of the \lcdm\ model is expressed in terms of the parameterized post-Newtonian (PPN) parameter $\ga_{PPN}$ \cite{Linder:2007hg,Will:2005va,Anton:2025zer}.
It is worth noting the growth index values in two different $f(R)$ gravity models to observe this scale dependence: The growth index for the Starobinsky model \cite{Starobinsky:2007hu} is calculated in \cite{Gannouji:2008wt} and found to be $\ga (z=0) = 0.399$ for $\Omega_{m0} = 0.315$. According to \cite{Tsujikawa:2009ku}, many feasible $f(R)$ gravity models have a growth index value of $\ga = 0.4$. In contrast, the growth index \cite{Basilakos:2017rgc} in the Hu-Sawicki model \cite{Hu:2007nk,Basilakos:2013nfa} is determined to be $\ga (a=1) = 0.753$, which is calculated on a different scale.

In this work, we demonstrate that even though the \gdcdm\ model is based on $f(R)$ gravity, the growth of structure behaves much closer to the case of \lcdm . We emphasize that a cosmological model that obeys modified gravity does not necessarily need to be yet another parametrization of dark energy, but it may introduce a completely new relationship between expansion of the Universe and its energy content. The theoretical value of the growth index, as determined by Bayesian analyses with various datasets, is found to be $\ga (a=1) = 0.561$. This value is very close to the \lcdm\ value of $\ga (a=1) = 0.555$.
When it comes to structure growth, the prediction of the growth index of the \gdcdm\ model is more in line with the \lcdm\ and \wcdm\ models than with the f(R) gravity theories.

In the \gdcdm\ model, we had to utilize the extended gravitational growth framework \cite{Linder:2009kq,Linder:2013dga}, which is forced by the non-standard high redshift limit of the density contrast, $\delta_m \propto a^{(1-\delta)}$. Thus, the growth rate had to be defined by including the growth rate calibration parameter $f_\infty$, as $f(a) = f_\infty \hat{\Omega}_m^\gamma$ (\ref{finf}). We find that the growth rate at high redshift is less than that found in models based on general relativity: $f \ra f_\infty = (1-\delta) < 1$ in the \gdcdm\ model. As discussed in Section \ref{models}, suppressed growth in turn causes the linear ISW effect to be larger than general relativistic models. Whether the enhancement of the late ISW signal in the \gdcdm\ model quantitatively matches the observed one \cite{Granett:2008ju} needs to be analyzed in future work.

To deal with the $S_8$ tension, we also need to calculate the $\si_8$ parameter in the \gdcdm\ model. To assess whether the \gdcdm\ model has relevance to $S_8$ tension, analyses with two distinct datasets are performed:
one is the $f\si_8(z)$ dataset \cite{Skara:2019usd,Kazantzidis:2018rnb}, and the other is the $\si_8(z)$ dataset \cite{Piccirilli:2024xgo,Oliveira:2025huk}. These datasets contain 35 and 14 data points, respectively. We combine these datasets with the DESI2 and DESY5 cosmological datasets to constrain the cosmological parameters of the \gdcdm\ model with smaller $1\si$ confidence regions. From analyses with these datasets we concluded that the \gdcdm\ model is indeed a competitive model to resolve the $S_8$ tension. 

The growth of structure is as important as expansion of the Universe in distinguishing scientific models that bring more sound explanations to the cosmological phenomena. We plan to analyze our model with further datasets, especially high redshift datasets to appraise its full potential, and to determine whether it can resolve the important tensions in theoretical cosmology.

\section*{Acknowledgments} 

This work is finished while C.D. is visiting A.P.C., Universit\'e Paris Cit\'e. C.D. thanks Maria Cristina Volpe for hospitality. Authors thank Eric V. Linder for critical comments on the manuscript and very helpful discussions, especially suggesting the extended gravitational growth framework. F.\c{S}.D. and S.S.B. also thank Daniel Scolnic for helpful discussions on the DESY5 dataset. F.\c{S}.D. is also supported by T\"UB\.{I}TAK-B\.{I}DEB 2211-A Domestic Ph.D. Scholarship.


\appendix

\section{Data tables \label{appdata}}

\begin{table}[hbt!]
\centering
\def\arraystretch{1.2}
\begin{tabular}{|c|c|c|c|}
\hline 
$z$ & $f(z)$ & Survey &  Cosmological tracer \\
\hline 
$0.013$ & $0.56\pm0.07$ & ALFALFA \cite{Avila:2021dqv} &HI extragalactic sources  \\
$0.15$ & $0.49\pm0.14$ & 2dFGRS \cite{Hawkins:2002sg,Guzzo:2008ac} & galaxies  \\
$0.18$ & $0.49\pm0.12$ & GAMA \cite{Blake:2013nif} & multiple-tracer: blue \& red galaxies  \\
$0.22$ & $0.60\pm0.10$ & WiggleZ \cite{Blake:2011rj} & galaxies  \\
$0.35$ & $0.70\pm0.18$ & SDSS \cite{SDSS:2006lmn} & luminous red galaxies (LRG)  \\
$0.38$ & $0.66\pm0.09$ & GAMA \cite{Blake:2013nif} &multiple-tracer: blue \& red galaxies \\
$0.41$ & $0.70\pm0.07$ & WiggleZ \cite{Blake:2011rj} & galaxies \\
$0.55$ & $0.75\pm0.18$ & 2SLAQ \cite{Ross:2006me} & LRG \& quasars  \\
$0.60$ & $0.73\pm0.07$ & WiggleZ \cite{Blake:2011rj} & galaxies \\
$0.77$ & $0.91\pm0.36$ & VIMOS-VLT Deep Survey \cite{Guzzo:2008ac} & faint galaxies  \\
$1.40$ & $0.90\pm0.24$ & 2QZ \& 2SLAQ \cite{daAngela:2006mf} & quasars  \\
\hline
\end{tabular} 
\caption{Direct measurements of the growth rate $f(z)$, independent of $\sigma_8$, as compiled in \cite{Avila:2022xad}.} 
\label{fzdata}
\end{table}

\begin{table}[h]
\centering
\def\arraystretch{1.2}
\begin{tabular}{|c|c|c|l|}
\hline
$z$ & $\sigma_8(z)$ & References \\
\hline
0.24 & $0.67 \pm 0.04$ & \cite{Garcia-Garcia:2021unp} \\
0.47 & $0.58 \pm 0.04$ & \cite{Kilo-DegreeSurvey:2023gfr} \\
0.53 & $0.59 \pm 0.03$ & \cite{Garcia-Garcia:2021unp} \\
0.60 & $0.59 \pm 0.02$ & \cite{ACT:2023oei} \\
0.63 & $0.53 \pm 0.04$ & \cite{Kilo-DegreeSurvey:2023gfr} \\
0.69 & $0.66 \pm 0.10$ & \cite{Piccirilli:2024xgo} \\
0.80 & $0.47 \pm 0.04$ & \cite{Kilo-DegreeSurvey:2023gfr} \\
0.83 & $0.58 \pm 0.04$ & \cite{Garcia-Garcia:2021unp} \\
0.92 & $0.44 \pm 0.06$ & \cite{Kilo-DegreeSurvey:2023gfr} \\
1.10 & $0.48 \pm 0.01$ & \cite{ACT:2023oei} \\
1.50 & $0.46 \pm 0.05$ & \cite{Garcia-Garcia:2021unp} \\
1.59 & $0.39 \pm 0.06$ & \cite{Piccirilli:2024xgo} \\
2.72 & $0.22 \pm 0.06$ & \cite{Piccirilli:2024xgo} \\
3.80 & $0.12 \pm 0.06$ & \cite{Miyatake:2021qjr} \\
\hline
\end{tabular}
\caption{$\sigma_8(z)$ data as compiled in \cite{Piccirilli:2024xgo}. This table includes 14 data points of Table 1 of \cite{Oliveira:2025huk}.}
\label{s8data}
\end{table}

\begin{table}[hbt!]
\centering
\def\arraystretch{1.2}
\resizebox{\textwidth}{!}{%
\begin{tabular}{|c|c|c|c|c|c|c|}
\hline
Dataset & $z$ & $f\sigma_8(z)$ & Fiducial Cosmology ($\Omega_{m0}$) 
& $q_{\scalebox{.9}{$\scriptscriptstyle \gamma\delta \mathrm{CDM}$}}$ 
& $q_{\scalebox{.9}{$\scriptscriptstyle 0.299$}}$ 
& $q_{\scalebox{.9}{$\scriptscriptstyle 0.315$}}$\\
\hline
2MRS \cite{Davis:2010sw,Hudson:2012gt} & 0.02 & $0.314 \pm 0.048$ & $0.266$ & 1.0016 & 1.0005 & 1.0007\\ 
SDSS-LRG-200 \cite{Samushia:2011cs} & 0.25 & $0.3512 \pm 0.058$ & $0.276$ & 1.0138 & 1.0044 & 1.0074\\ 
WiggleZ \cite{Blake:2012pj} & 0.44 & $0.413 \pm 0.080$ & $0.27$ & 1.0208 & 1.0093 & 1.0143\\ 
WiggleZ \cite{Blake:2012pj} & 0.60 & $0.390 \pm 0.063$ & $0.27$ & 1.0231 & 1.0120 & 1.0184\\ 
WiggleZ \cite{Blake:2012pj} & 0.73 & $0.437 \pm 0.072$ & $0.27$ & 1.0236 & 1.0137 & 1.0210\\ 
GAMA \cite{Blake:2013nif} & 0.18 & $0.360 \pm 0.090$ & $0.27$ & 1.0117 & 1.0040 & 1.0062\\ 
SDSS-MGS \cite{Howlett:2014opa} & 0.15 & $0.490 \pm 0.145$ & $0.31$ & 1.0055 & 0.9988 & 1.0006\\ 
SDSS-veloc \cite{Feix:2015dla,SDSS:2003tbn} & 0.10 & $0.370 \pm 0.130$ & $0.3$ & 1.0048 & 0.9999 & 1.0011\\ 
FastSound \cite{Okumura:2015lvp,WMAP:2012nax} & 1.40 & $0.482 \pm 0.116$ & $0.27$ & 1.0183 & 1.0184 & 1.0280\\ 
BOSS DR12 \cite{BOSS:2016wmc} & 0.38 & $0.497 \pm 0.045$ & $0.31$ & 1.0081 & 0.9970 & 1.0014\\ 
BOSS DR12 \cite{BOSS:2016wmc} & 0.51 & $0.458 \pm 0.038$ & $0.31$ & 1.0075 & 0.9962 & 1.0017\\ 
BOSS DR12 \cite{BOSS:2016wmc} & 0.61 & $0.436 \pm 0.034$ & $0.31$ & 1.0064 & 0.9956 & 1.0020\\ 
VIPERS v7 \cite{Wilson:2016ggz} & 1.05 & $0.280 \pm 0.080$ & $0.308$ & 1.0004 & 0.9951 & 1.0037\\ 
BOSS LOWZ \cite{Gil-Marin:2016wya} & 0.32 & $0.427 \pm 0.056$ & $0.31$ & 1.0079 & 0.9974 & 1.0012\\ 
VIPERS \cite{Hawken:2016qcy} & 0.727 & $0.296 \pm 0.0765$ & $0.31$ & 1.0048 & 0.9951 & 1.0022\\ 
6dFGS+SnIa \cite{Huterer:2016uyq} & 0.02 & $0.428 \pm 0.0465$ & $0.3$ & 1.0011 & 1.0000 & 1.0002\\ 
2MTF \cite{Howlett:2017asq} & 0.001 & $0.505 \pm 0.085$ & $0.3121$ & 1.0001 & 1.0000 & 1.0000\\ 
BOSS DR12 \cite{Wang:2017wia} & 0.31 & $0.384 \pm 0.083$ & $0.307$ & 1.0086 & 0.9982 & 1.0018\\ 
BOSS DR12 \cite{Wang:2017wia} & 0.36 & $0.409 \pm 0.098$ & $0.307$ & 1.0088 & 0.9979 & 1.0021\\ 
BOSS DR12 \cite{Wang:2017wia} & 0.40 & $0.461 \pm 0.086$ & $0.307$ & 1.0089 & 0.9977 & 1.0023\\ 
BOSS DR12 \cite{Wang:2017wia} & 0.44 & $0.426 \pm 0.062$ & $0.307$ & 1.0089 & 0.9975 & 1.0025\\ 
BOSS DR12 \cite{Wang:2017wia} & 0.48 & $0.458 \pm 0.063$ & $0.307$ & 1.0087 & 0.9973 & 1.0026\\ 
BOSS DR12 \cite{Wang:2017wia} & 0.52 & $0.483 \pm 0.075$ & $0.307$ & 1.0085 & 0.9972 & 1.0028\\ 
BOSS DR12 \cite{Wang:2017wia} & 0.56 & $0.472 \pm 0.063$ & $0.307$ & 1.0082 & 0.9970 & 1.0030\\ 
BOSS DR12 \cite{Wang:2017wia} & 0.59 & $0.452 \pm 0.061$ & $0.307$ & 1.0079 & 0.9969 & 1.0031\\ 
BOSS DR12 \cite{Wang:2017wia} & 0.64 & $0.379 \pm 0.054$ & $0.307$ & 1.0073 & 0.9967 & 1.0033\\ 
SDSS-IV \cite{eBOSS:2018yfg} & 0.978 & $0.379 \pm 0.176$ & $0.31$ & 1.0006 & 0.9943 & 1.0026\\ 
SDSS-IV \cite{eBOSS:2018yfg} & 1.23 & $0.385 \pm 0.099$ & $0.31$ & 0.9962 & 0.9938 & 1.0028\\ 
SDSS-IV \cite{eBOSS:2018yfg} & 1.526 & $0.342 \pm 0.070$ & $0.31$ & 0.9916 & 0.9935 & 1.0029\\ 
SDSS-IV \cite{eBOSS:2018yfg} & 1.944 & $0.364 \pm 0.106$ &  $0.31$& 0.9862 & 0.9933 & 1.0030\\ 
VIPERS PDR2 \cite{Mohammad:2018mdy} & 0.60 & $0.49 \pm 0.12$ & $ 0.31$ & 1.0066 & 0.9957 & 1.0019\\ 
VIPERS PDR2 \cite{Mohammad:2018mdy} & 0.86 & $0.46 \pm 0.09$ &  $ 0.31$  & 1.0026  & 0.9946 & 1.0024\\ 
BOSS DR12 voids \cite{Nadathur:2019mct} & 0.57 & $0.501 \pm 0.051$ & $ 0.307$ & 1.0081 & 0.9970 & 1.0030\\ 
2MTF 6dFGSv \cite{Qin:2019axr} & 0.03 & $0.404 \pm 0.0815$ & $0.3121$ & 1.0014 & 0.9997 & 1.0001\\ 
SDSS-IV \cite{eBOSS:2019gbd} & 0.72 & $0.454 \pm 0.139$ & $0.31$ & 1.0049 & 0.9951 & 1.0022\\ 
\hline
\end{tabular}
}
\caption{$f\sigma_8 (z)$ data as compiled in \cite{Skara:2019usd,Kazantzidis:2018rnb}. These are subset of data points with less correlation, which are indicated in bold font in Table VI of appendix B of \cite{Skara:2019usd}. Last three columns include the Alcock-Paczynsk correction factor $q$ \cite{Skara:2019usd,Kazantzidis:2018rnb,Nesseris:2017vor,Macaulay:2013swa} for the \gdcdm\ model, the \lcdm\ model with $\Omega_{m0} = 0.299$, and the \lcdm\ model with $\Omega_{m0} = 0.315$.}
\label{fsigma8}
\end{table}

\end{document}